\documentclass[preprint]{elsarticle}
\usepackage{amssymb}
\usepackage{algorithm2e}
\usepackage{amsthm}
\usepackage{amsmath}

\usepackage[T1]{fontenc} 
\usepackage{booktabs}
\usepackage{hyperref}
\usepackage{caption}
\usepackage{xcolor}
\usepackage{float}
\usepackage{lineno}
\usepackage{pifont}
\usepackage{booktabs}
\biboptions{numbers,sort&compress}

\makeatletter
\def\ps@pprintTitle{%
  \let\@oddhead\@empty
  \let\@evenhead\@empty
  \let\@oddfoot\@empty
  \let\@evenfoot\@oddfoot
}
\makeatother

\AtBeginDocument{}
\DeclareMathOperator*{\argmin}{argmin} 

\begin{document}

\captionsetup[figure]{labelfont={bf},labelformat={default},labelsep=period,name={Fig.}}
\begin{frontmatter}

\title{Bayesian approach for uncertainty quantification of hybrid spectral unmixing in $\gamma$-ray spectrometry}

\author[inst1]{Dinh Triem Phan\corref{cor1}}
\cortext[cor1]{Corresponding author}

\ead{dinh-triem.phan2@cea.fr}
\affiliation[inst1]{organization={Université Paris-Saclay, CEA, List, Laboratoire national Henri Becquerel (LNE-LNHB)},
            city={Palaiseau},
            postcode={91120}, 
            country={France}}

\author[inst2]{Jérôme Bobin}
\author[inst1]{Cheick Thiam}
\author[inst1]{Christophe Bobin}
\affiliation[inst2]{organization={IRFU, CEA, Universite Paris-Saclay},
            city={Gif-sur-Yvette},
            postcode={91191}, 
            country={France}}

\begin{abstract}
Identifying and quantifying $\gamma$-emitting radionuclides, considering spectral deformation from $\gamma$-interactions in radioactive source surroundings, present a significant challenge in $\gamma$-ray spectrometry. In that context, a hybrid machine learning method has been previously proposed to jointly estimate the counting and spectral signatures of $\gamma$-emitters under conditions of spectral variability. This paper addresses the uncertainty quantification of the estimators {(\it i.e.,} the counting and the variable $\lambda$ which characterizes the spectral signatures) obtained by this spectral unmixing algorithm. The focus is on the coverage interval, as defined by the GUM, which corresponds closely to a credible interval in the Bayesian framework. Given the inverse problem and the constraints associated with spectral deformation, two Bayesian methods - Laplace approximation and Markov Chain Monte Carlo - have been developed for uncertainty quantification to ensure robust decision-making. The Laplace approximation technique approximates the posterior distribution by a Gaussian distribution, while the Markov Chain Monte Carlo technique samples the posterior distribution. This study evaluates these two methods in terms of precision of coverage interval based on repeated Monte Carlo samples using the long-run success rate. Numerical experiments show that both methods yield similar results close to the expected success rate of 95.4$\%$ when constraints related to spectral signatures deformation and counting are inactive. However, when constraints are active or the background counting significantly dominates other radionuclides, the Laplace approximation method deviates from the expected long-run success rate due to the non-Gaussian posterior distribution. In such cases, the Markov Chain Monte Carlo method still provides robust results.
\end{abstract}

\end{frontmatter}

\newpage
\section{Introduction}

Gamma-ray spectrometry is a well-established technique \cite{gilmore2008practical} for the identification and quantification of $\gamma$-emitting radionuclides in various nuclear applications. In radionuclide laboratories, the analysis of $\gamma$-spectra for metrology purposes is generally carried out by experts using full absorption peaks related to energies depending on radionuclide decay schemes. The procedure is traditionally based on peak fitting techniques assuming Gaussian statistics— a valid approximation when counting is high, as Poisson statistics approach a Gaussian distribution. Reliable results are also important in the case of {\it in situ} $\gamma$-ray spectrometry for applications such as environmental measurements ({\it e.g.}, following a radiological or nuclear accident), radiological characterization for decommissioning of nuclear installations \cite{ac024598661998radiological} or security screening for illegal nuclear material trafficking \cite{euro2007combating}. In those cases, the need for specific algorithms yielding good performances is important for fast automatic identification and quantification of $\gamma$-emitting radionuclides, particularly at low statistics. 

In that context, full-spectrum statistical methods based on spectral unmixing \cite{andre2021metrological,xu2020sparse} make it possible to meet metrological constraints such as robust decision-making and low-bias estimation of counting related to identified radionuclides. This method decomposes an experimental $\gamma$-spectrum - represented as a vector $y \in \mathbb{R}^{M}$, where $M$ is the number of channels - into individual spectral signatures which correspond to the detector’s characteristic response to $\gamma$-photon emissions from various radionuclides.

In this framework, the observed spectrum $y$ can be modeled by a Poisson distribution of $Xa$: $y \sim \mathcal{P}(Xa)$ where $X \in \mathbb{R}^{M\times N} $ is a matrix of the normalized spectral signature of all $N$ radionuclides, including the natural background (Bkg). The spectral signatures can be obtained through simulations or experiments and are normalized by dividing each spectrum by its total number of counts. The vector $a \in \mathbb{R} ^{N}$ contains the counting of each radionuclide. A regularized Maximum Likelihood Estimation (MLE) is then used to estimate the counting of all radionuclides. This method has been applied to NaI(Tl) and HPGe detectors and demonstrated to give better results in terms of activity estimation and decision thresholds than traditional methods \cite{xu2020sparse,malfrait2023spectral, malfrait2023online, mano2024algorithm}.  

In the reference \cite{andre2021metrological}, a metrological approach to this statistical technique, including uncertainty quantification, has been proposed for the case where spectral signatures are fixed corresponding to well-defined measurement conditions. However, {\it in situ} measurements generally take place in complex measurement conditions, leading to variability in spectral signatures due to physical phenomena such as attenuation, Compton scattering occurring outside the detector ({\it e.g.,} shielding materials). In this case, the matrix $X$ depends on the measurement conditions and is not known in advance. To deal with this issue, a hybrid machine learning (ML) unmixing algorithm SEMSUN was developed in a previous work \cite{phan2024hybrid}, which combines a pre-trained ML model based on an autoencoder architecture that captures spectral deformation and a regularized MLE to determine both the counting of the radionuclides and their spectral signatures.

To improve the reliability of the results given by SEMSUN, it is important to assess the uncertainty of the measurands: the estimated counting and the estimated variable $\lambda$ which characterizes the spectral signatures. The uncertainty assessment is also implemented for $\lambda$, as it reflects the deformation of spectral signatures and can be used to infer physical properties, such as material thickness or the absolute total efficiency of the detection system. Radionuclide activity can then be calculated from the counting and the absolute total efficiency. However, to properly account for activity, other sources of uncertainty must be considered, including $\gamma$-ray emission intensities, corrections for the dead time, the pre-trained ML model, and the absolute total efficiency. These factors will be explored in more detail in future research. For the scope of this work, the emphasis is on counting uncertainty, which is mainly driven by statistical noise.

In the Guide to the Expression of Uncertainty in Measurement (GUM) \cite{bipm2008evaluation} framework, measurement uncertainty can be expressed using various manners such as the standard uncertainty and the coverage interval \cite{bipm2008evaluation}. This work focuses on the coverage interval (CI), which corresponds closely to a credible interval in the Bayesian framework \cite{bipm2008evaluation1, stoudt2021coverage, elster2014bayesian,giaquinto2016examples}. Therefore, this work investigates a Bayesian approach that includes two distinct approximation and computation methods: Laplace approximation (LA) and Markov Chain Monte Carlo (MCMC) \cite{klauenberg2016markov}. The LA technique approximates the posterior distribution by a Gaussian distribution, while the MCMC technique samples the posterior distribution. This study evaluates these methods in terms of precision of the coverage interval based on repeated Monte Carlo samples using the long-run success rate (LRSR) \cite{hall2008evaluating,willink2010probability}. A user guide is also introduced to efficiently apply the uncertainty quantification method for spectral unmixing in $\gamma$-ray spectrometry. \\

The article is organized as follows:
\begin{itemize}
\item {Section 2 provides a brief introduction of spectral unmixing in $\gamma$-ray spectrometry.}
\item {Two Bayesian methods, LA and MCMC, for uncertainty calculation of SEMSUN estimator in the case of spectral variability are presented in Section 3. }
\item {Section 4 details the methodology used to evaluate the coverage interval through LRSR, and the results of both techniques are presented in Section 5.}

\end{itemize}

\section{Brief introduction of spectral unmixing in $\gamma$-ray spectrometry}
In the spectral unmixing approach, an observed spectrum $y$ follows the Poisson distribution of $Xa$: $y \sim \mathcal{P}(Xa)$ where $X=[X_{Bkg},X_2,..,X_N] \in \mathbb{R}^{M\times N}$ is the matrix with the column $j$ being the spectral signature of radionuclide $j$ (including Bkg) and $a \in \mathbb{R} ^{N}$ is the vector containing the counting of each radionuclide. Based on the statistical independence of the measurement channels, the joint probability for the likelihood can be described as: 
\begin{equation}
p(y|X,a)=\prod_{m=1}^M \frac{(Xa)_m^{y_m} e^{ (Xa)_m}}{y_m!}
\label{eq:poisson}
\end{equation}
Therefore, the cost function to be minimized can be defined as the negative logarithm of the likelihood which has the following formula:
\begin{equation}
L(y,X,a)=\sum_{m=1}^M ((Xa)_m - y_m log ((Xa)_m))
\label{eq:divergence}
\end{equation}

When the spectral signatures $X$ are fixed and known, the problem boils down to estimating the non-negative counting vector $a$ from the observed spectrum $y$ which minimizes the cost function $L$:
\begin{equation} 
         \hat{a} =\argmin_{a\geq 0}  L(y,X,a) 
\label{eq:pb1}
\end{equation}
In the case of spectral variability, the matrix $X$ depends on the measurement conditions and is not known in advance. For a proof-of-concept purpose in \cite{phan2024hybrid}, spectral signatures of a NaI(Tl) detector are mainly deformed due to attenuation and Compton scattering occurring outside the detector when the radioactive source is placed in a steel sphere of varying thickness. In this case, $X_{j \ge 2}$ can be modeled by a non-linear function of a latent variable $\lambda$ learned by a particular Machine learning model (coined Interpolating AutoEncoder (IAE) \cite{bobin2023autoencoder, phan2024hybrid}) using Geant4 \cite{agostinelli2003geant4} simulations: $X_{j \ge 2} \approx f(\lambda) \, ; \lambda \in [0,1] $ (Bkg ($X_1$) is supposed to be known). Mathematically, the spectral unmixing problem can be recast to estimate the spectral signatures $X$ (or the latent variable $\lambda$) and the counting vector $a$ that minimizes the cost function $L$:
\begin{equation} 
         \hat{\lambda},\hat{a} =\argmin_{\lambda \in [0,1],a\geq 0}  L(y, f(\lambda),a)    ;  \quad  \hat{X}_{j \ge 2}= f (\hat{\lambda} ) 
\label{eq:pb2}
\end{equation}
The semi-blind spectral unmixing (SEMSUN) \cite{phan2024hybrid} was developed to solve this above optimization problem. This algorithm has been demonstrated to provide robust identification and accurate quantification of $\gamma$-emitting radionuclides, particularly under low-statistics conditions \cite{phan2025automatic}.

This article addresses the uncertainty assessment of the measurands $\theta=(a,\lambda)$ obtained with the hybrid algorithm SEMSUN when using a single measured spectrum $y$ ({\it i.e.,} the uncertainty quantification is not based on repeated measurements). The focus is on the coverage interval (CI), which is defined as an interval containing the value of a quantity with a stated probability using the available data \cite{bipm2008evaluation1}. Given the multivariate nature of the measurand in this study, Supplement 2 to the GUM, which relies on the Monte Carlo method, is the most appropriate. However, in this case, the problem involves a generalized linear model, where the measurement model is not analytically available. It can only be calculated through the SEMSUN algorithm $\theta = SEMSUN(y)$. Employing Monte Carlo methods with SEMSUN is computationally intensive and time-consuming, making it not applicable for fast decision-making. As an alternative, the Bayesian framework is particularly well-suited for this problem since a credible interval in the Bayesian framework aligns closely with the definition of the coverage interval \cite{bipm2008evaluation1, stoudt2021coverage, elster2014bayesian,giaquinto2016examples}. Additionally, it naturally takes into account constraints on the measurands through the prior distributions.

\section{Uncertainty quantification for spectral unmixing in $\gamma$-ray spectrometry}

\subsection{Spectral unmixing in Bayesian framework}\label{sec:bay}
In Bayesian statistics, the process begins with selecting a prior distribution $p(\theta)$. When measuring a new spectrum $y$, the Bayes' theorem is used to update this prior belief to form a posterior distribution $ p(\theta|y)$:
\begin{equation} 
       p(\theta|y)=p(y|\theta) p(\theta)/p(y)
\label{eq:baye1}
\end{equation}
where $p(y|\theta)$ represents the likelihood function and $p(y)$ is the evidence. Assuming uniform priors for the parameters $\theta=(a,\lambda)$: $a_i \sim U(0,\infty)$, $\lambda \sim U(0,1)$, the prior distributions are constant within their respective domains:  $p(\lambda)=1 \; for \; \lambda \in [0,1]$ and $p(a_i)=1 \; for \; a_i \in [0,\infty]$. Given these priors, the posterior distribution simplifies as follows:
\begin{equation} 
        p(\theta|y)  \propto p(y|\theta) p(\theta)  \propto p(y|\theta) p(\lambda) \prod_{i=1} ^N p(a_i)
\label{eq:fisher}
\end{equation}
Therefore, the regularized MLE obtained by SEMSUN in \ref{eq:pb2} is equal to the maximum a posteriori estimate (MAP) in the Bayesian framework.

In Bayesian inference, the posterior distribution is essential for quantifying uncertainty. Several methods are available for calculating this distribution such as conjugate priors, Laplace approximation, Variational Bayesian methods or expectation propagation \cite{van2000asymptotic, beal2003variational}. These methods require specific assumptions about the prior distribution or the form of the posterior distribution. In addition to these methods, MCMC methods have become increasingly useful in recent years thanks to advances in computing power. MCMC is a powerful technique for sampling complex posterior distributions without specifying their exact form in advance by generating a chain of dependent samples. This work investigates the LA method, valued for its simplicity and computational efficiency under the assumption of Gaussian posterior and an MCMC method for dealing with the uncertainty problem.

\subsection{Laplace approximation (LA)}\label{sec:la}

The Laplace approximation method \cite{beal2003variational,kass1995bayes} proposes to approximate the posterior distribution by applying a Taylor expansion to the logarithm of the posterior density. This yields a Gaussian distribution centered at the MAP estimate, with a covariance matrix given by the inverse of the Hessian of the log-posterior. Given the use of uniform priors in this study, the MAP coincides with the MLE, and this covariance matrix equals the inverse of Fisher information matrix. It should be noted that, in this case, the uncertainty resulting from the LA is equivalent to that obtained via the Fisher information matrix in a frequentist framework.

The Fisher matrix \cite{fisher1922mathematical,fisher1925theory} is defined as the expectation of the second derivative of the negative log-likelihood:
 $$I(\theta)=E_\theta\left[\frac{\partial^2}{\partial \theta^2} L(\theta ; y)\right] = E_\theta\left[-\frac{\partial^2}{\partial \theta^2} \ln p(y|\theta )\right] $$
To compute the Fisher information matrix, derivatives of the cost function $L$ with respect to the parameter $\theta$ are needed. When the spectral signatures are known ($\theta=(a)$), an analytical formula for the derivatives exists \cite{andre2021metrological}, enabling the calculation of $I(\theta)$ as: 
\begin{equation} 
         I(\theta)_{ij} = \sum_{m=1}^M \frac{X_{mi} X_ {mj}}{(Xa)_m}
\label{eq:fisher}
\end{equation}
However, for spectral variability ($\theta=(a,\lambda)$), analytical derivatives with respect to $\lambda$ are unavailable since the function $f$ is modeled by the IAE neural network. To address this, PyTorch's autograd \cite{NEURIPS2019_9015} function can be used to approximate this calculation.

Using this method, a coverage interval $(1-\alpha)100\%$ CI for each parameter $\theta_i$ can be defined as follows: $CI(\theta_i)=[\hat \theta_i \pm \sigma _{\theta_i} z_{\alpha/2}]$ where $ \hat \theta$ is the solution provided by SEMSUN, $\sigma _{\theta_i}$ is the standard deviation derived from the Fisher information matrix and $ z_{\alpha/2}$ is the inverse of the cumulative distribution function of the standard normal distribution.

This method assumes that the posterior distribution follows a Gaussian distribution. However, constraints on parameters ($\lambda \in [0,1]$ and $a_i \geq 0$) may lead the distribution not to follow a Gaussian distribution in some cases. As a result, the LA method may yield imprecise CI when these constraints significantly affect the distribution. To address this issue, a simple criterion can be established for each observed spectrum to assess the applicability of this method. This criterion evaluates whether the estimator distributions are a Gaussian or a truncated Gaussian by checking the activation of constraints. A straightforward method involves verifying whether the estimator plus or minus three standard deviations, which is a good approximation for a Gaussian distribution, satisfies the constraints. For instance, if the estimated counting of a radionuclide minus three standard deviations remains greater than zero, it is almost certain that the non-negativity constraint is not active. The criterion can be formalized as follows : $\hat \lambda - 3 \sigma _{\hat \lambda} \geq 0 \enspace   ;  \enspace    \hat \lambda + 3   \sigma _ {\hat \lambda} \leq 1 \enspace   ;  \enspace   \hat a_i -3  \sigma _{\hat a_i} \geq 0 \enspace  \forall i $
where $\hat \lambda$, $\hat a$ are the estimators obtained by SEMSUN.

It is noteworthy to highlight that even without constraints on $\lambda$ and $a$, there is no theoretical guarantee that the posterior distribution can be approximated by a Gaussian distribution since it is a generalized linear model combined with an ML component. This can only be empirically demonstrated when the constraints on $\theta$ are inactive.

\subsection{ Markov Chain Monte Carlo (MCMC)}

Under the Bayesian approach, there are various CIs that can provide the same nominal coverage probability \cite{van2000asymptotic}. A common method is the equal-tailed interval, where the probability of the parameter falling below the interval equals the probability of it exceeding the interval. Another option is the Highest Posterior Density (HPD) interval, identified as the narrowest interval of all CIs. The HPD interval contains the MAP, making it suitable for this study. When the posterior distribution is available, a large number of samples $T$ for the parameter $\theta=(a,\lambda)$ can be drawn from it, denoted as $\{\theta^{(1)},..,\theta^{(T)}\}$. {For each component $\theta_i$ (either $a_i$ or $\lambda$), the samples $\{ \theta^{(1)}_i,..,\theta^{(T)}_i\}$ are sorted as $\{ \overline \theta^{(1)}_i,..,\overline \theta^{(T)}_i\}$. Different CIs containing $(1-\alpha)\%$ of the samples are then formed as  $[\overline \theta^{(k)}_i,\overline \theta ^{(k+(1-\alpha) \times T-1)}_i]$. The HPD interval corresponds to the shortest of these candidate intervals.

Obtaining these posterior samples and constructing CIs requires efficient sampling techniques, for which MCMC methods are commonly employed. MCMC methods are powerful algorithms for estimating complex probability distributions by constructing a Markov chain that generates dependent samples. A classical MCMC technique is the Metropolis-Hastings (MH) algorithm \cite{metropolis1953equation}, which generates samples by proposing candidate states through a proposal distribution and accepting or rejecting them based on a probability ratio. The original MCMC algorithm is the MH with a Gaussian distribution as its proposal mechanism, which is generally called Random Walk Metropolis. Recently, Hamiltonian Monte Carlo (HMC) methods \cite{duane1987hybrid,mackay2003information} have been frequently used, especially for high-dimensional spaces. The HMC algorithm can be viewed as a specific MH algorithm that modifies the proposal mechanism using a random momentum vector and Hamiltonian dynamics. 

The No U-Turn Sampler (NUTS) \cite{hoffman2014no}, an extension of the HMC algorithm, is proven to efficiently sample complex distributions without manual tuning of the parameter. This study employs the MCMC with NUTS algorithm using the Pyro package \cite{bingham2019pyro}, which facilitates Bayesian modeling and ML in Python. In this setting, 300 samples during the burn-in period and 1000 posterior samples were selected afterward to run the MCMC algorithm. These posterior samples of $\theta$ were then used to determine the $(1 - \alpha)\%$ CI associated with $a$ and $\lambda$ based on the HPD interval described earlier.

\section{Numerical evaluation}
\subsection{Dataset}

The dataset for uncertainty assessment is the same as used in the reference \cite{phan2024hybrid}. Spectral deformations were simulated using Geant4 \cite{agostinelli2003geant4}, a Monte Carlo toolkit for simulating particle interactions with matter. A point source was placed in a steel sphere with different thicknesses, located 20 cm from the 3"×3" NaI(Tl) detector. In this case, the spectral signature of each radionuclide was deformed by the attenuation and Compton scattering effects. The dictionary of radionuclides included four radionuclides covering a large range of energies between 20 keV and 2 MeV:  $^{60}$Co,  $^{88}$Y, $^{99m}$Tc,  $^{137}$Cs. For each radionuclide, 96 spectral signatures were produced through Geant4 simulations, with the steel sphere thickness ranging from 0.001 mm to 30 mm. The NaI(Tl) detector's energy resolution was taken into account (6.5$\%$ at 662 keV). Each spectral signature is composed of 1024 channels, with a 2 keV bin width and a low-energy cut-off of 20 keV. The architecture and hyperparameters of the IAE remained the same as described in the reference \cite{phan2024hybrid}. A single IAE model was trained to represent the spectral signatures of all radionuclides, with the $\lambda$ dimension kept at one, using the same number of anchor points (see \cite{phan2024hybrid,bobin2023autoencoder} for more details). The SEMSUN-j version of the hybrid algorithm was used in this case. The spectral signatures of $^{60}$Co and $^{137}$Cs as a function of the thickness of steel sphere are shown in the \autoref{figure:spectre_X}. Additionally, \autoref{figure:example_spec} displays the spectral signatures of four radionuclides with a steel thickness of 20 mm, as well as an example of a simulated $\gamma$-spectrum at low statistics. The data and codes are available on: \url{https://github.com/triem1998/GamUncertainty}.
\begin{figure}[H]
    \centering
      \includegraphics[scale=0.18]{ 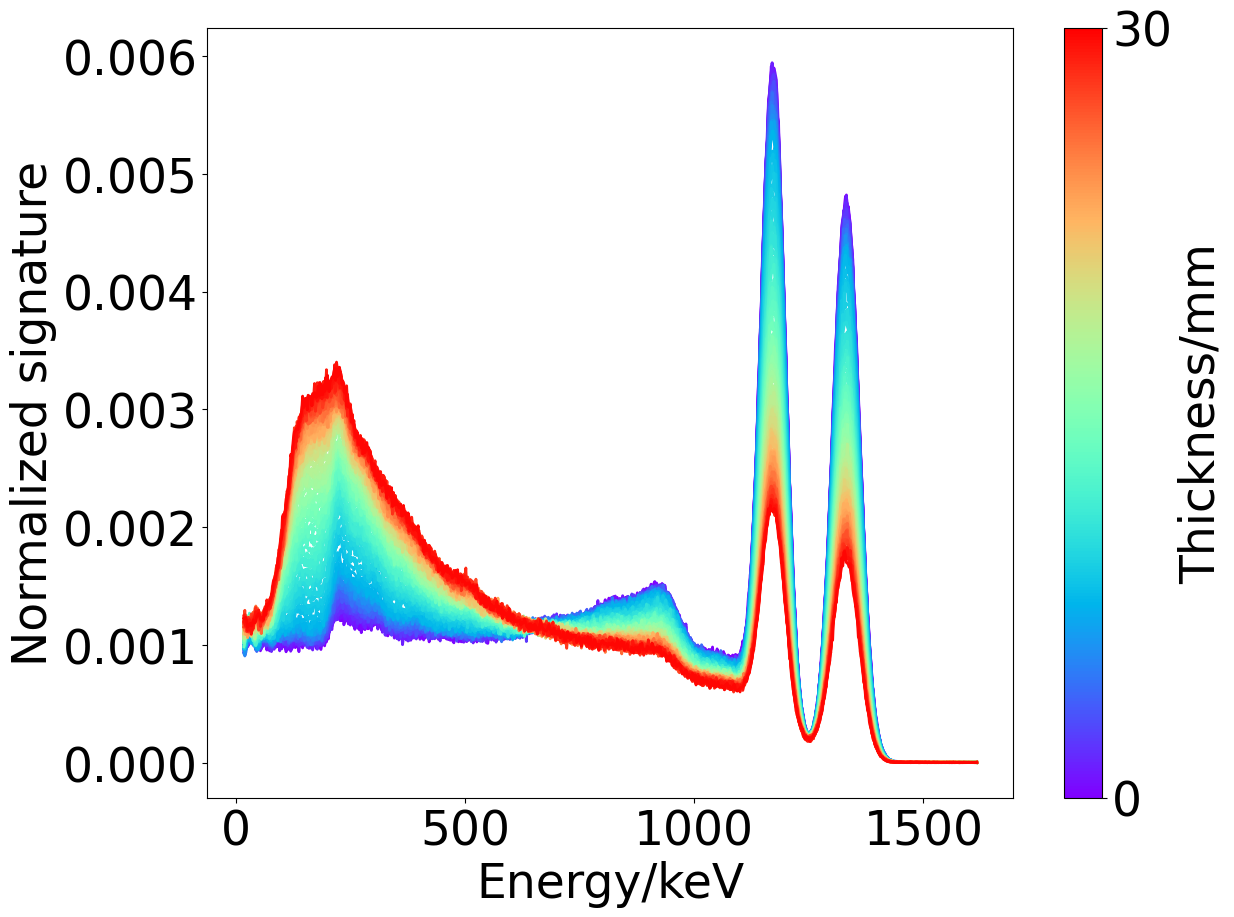}
        \includegraphics[scale=0.18]{ 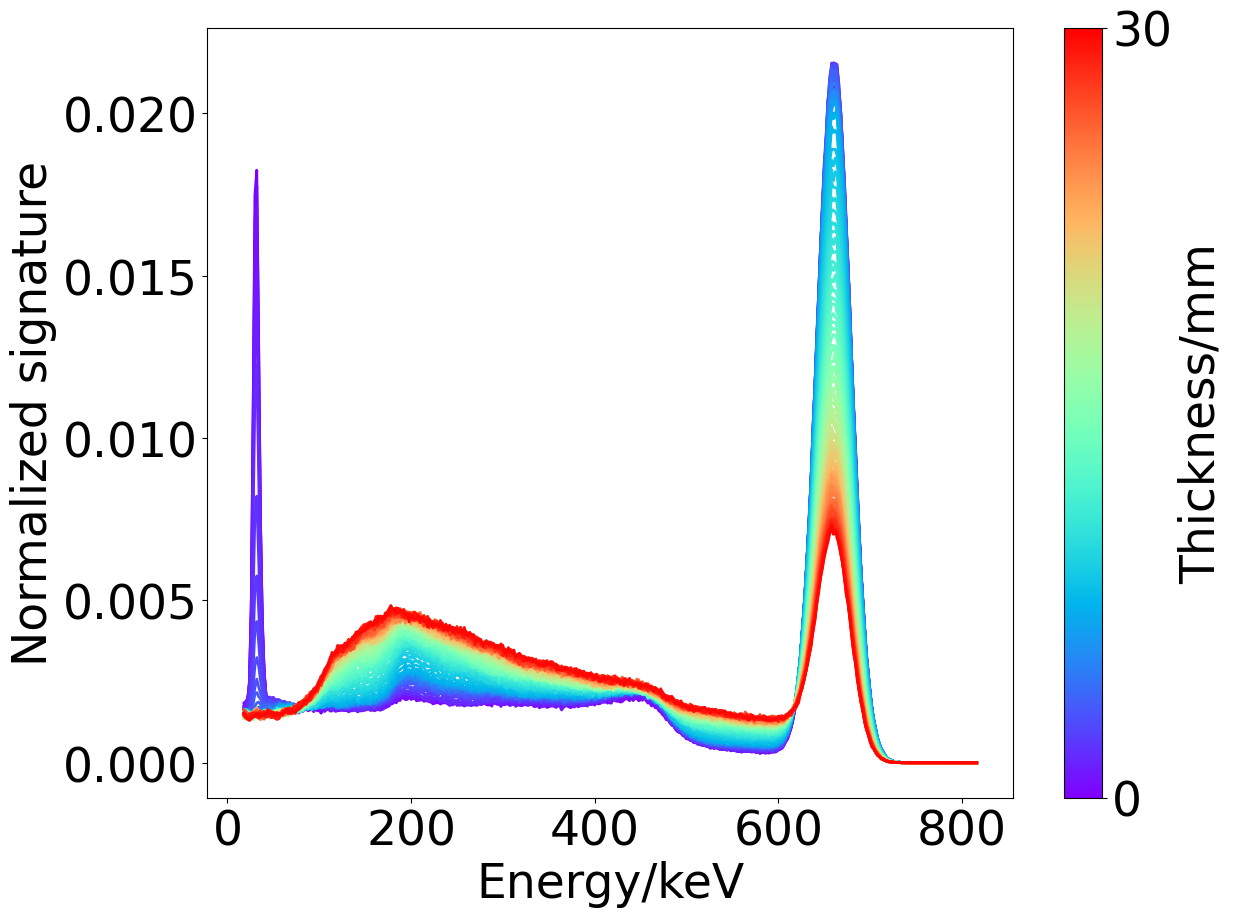}
    \caption{ Spectral signatures of $^{60}$Co (left) and $^{137}$Cs (right) as a function of steel thickness.  }
      \label{figure:spectre_X}
\end{figure}

\begin{figure}[H]
    \centering
      \includegraphics[scale=0.2]{ 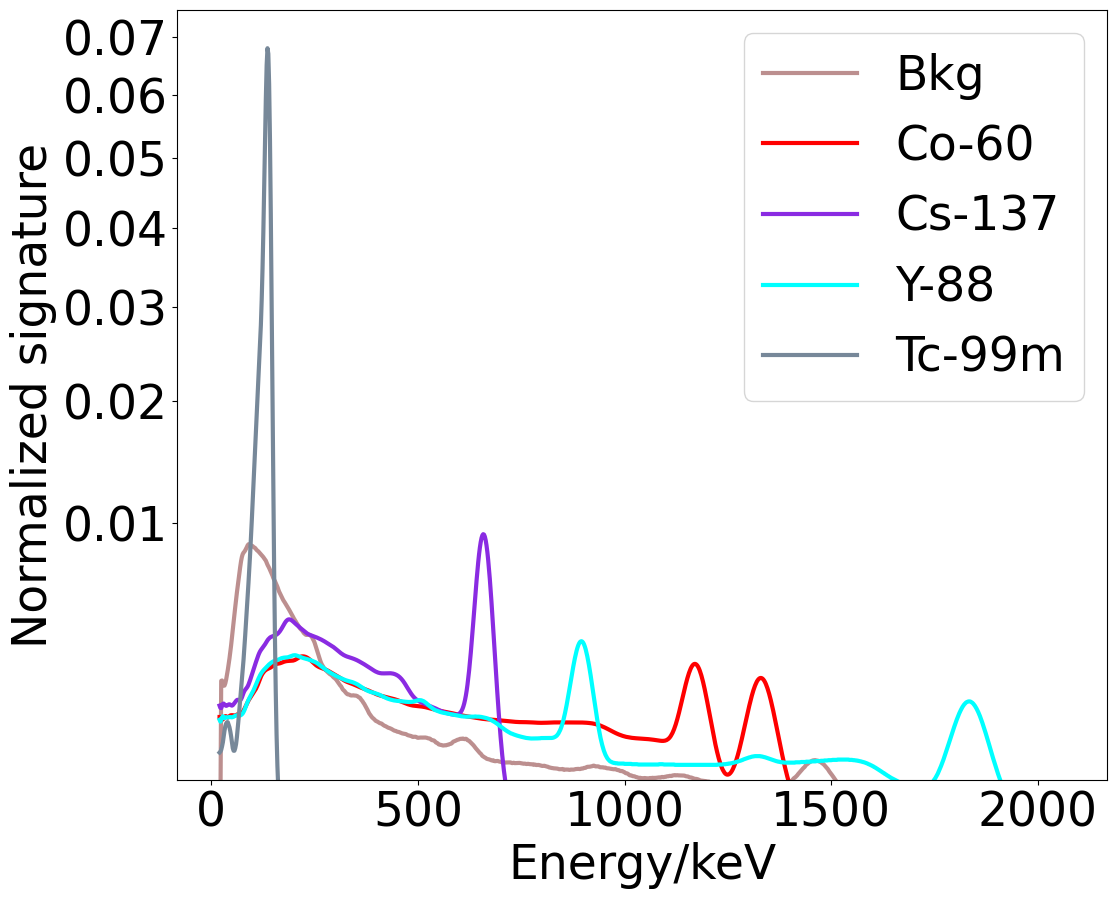}
        \includegraphics[scale=0.2]{ 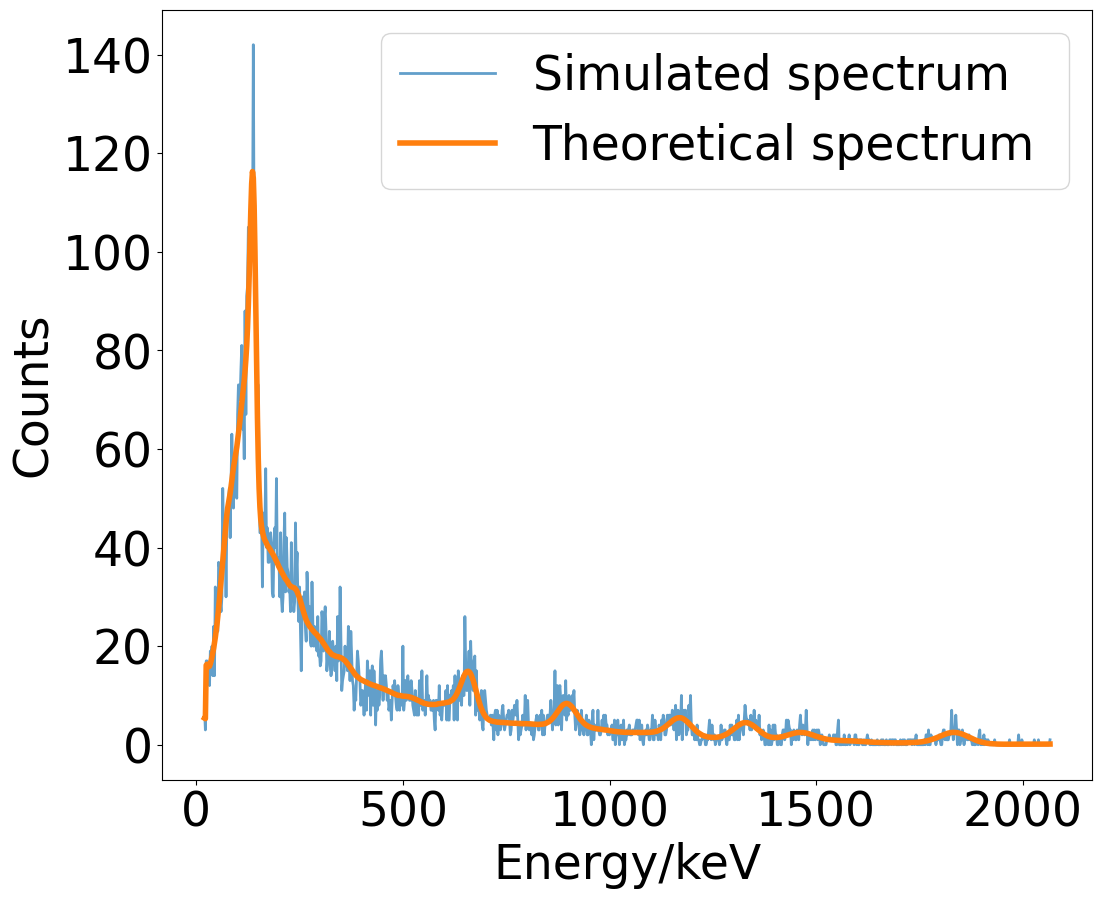}
    \caption{ Spectral signatures of 4 radionuclides with 20 mm steel thickness (left) and a simulated $\gamma$-spectrum of this mixture including $^{60}$Co,  $^{88}$Y, $^{99m}$Tc,  $^{137}$Cs and Bkg at low statistics (right). }
      \label{figure:example_spec}
\end{figure}

\subsection{Validation method of coverage interval for the spectral unmixing}\label{sec:eval_ci}
As discussed previously, the coverage interval defined in the GUM typically corresponds to the credible interval in the Bayesian framework. The validation of this interval presents significant challenges since it is a conditional probability on the observed data. The robustness of credible interval is mathematically guaranteed only in simple cases where the posterior distribution has an analytical formula. In more complex cases where the posterior distribution is approximated ({\it e.g.}, via MCMC), there is no straightforward way to empirically verify the coverage probability.

In metrology, the long-run success rate (LRSR) is introduced as a metric to evaluate the performance of a coverage interval through repeated sampling \cite{hall2008evaluating}. For instance, a procedure with a 95$\%$ LRSR is expected to produce coverage intervals that include the measurand in approximately 95$\%$ of independent measurements. Although the coverage probability of credible intervals is generally not equivalent to the LRSR, it is well established that poor Bayesian analyses are typically associated with very low LRSR values \cite{giaquinto2016examples}. Hence, in this work, LRSR is used to assess the performance of CI calculated by both the LA and MCMC methods.

To implement this validation, a true value of the measurand $\widetilde \theta$ is selected, and a large number of independent simulated measurements of $\gamma$-spectrum are computed with Monte Carlo (MC) simulations:

\begin{equation} 
        y^{(1)},..,y^{(L)} \sim \mathcal{P}(\widetilde X \widetilde a)
\label{eq:val}
\end{equation}
For each MC simulated spectrum $y^{(i)}$, a CI is calculated using both LA and MCMC methods. The LRSR is then determined by counting the number of times the CIs contain the true value and then dividing this count by the total number of simulations $(\sum_{i=1} ^N 1_{\widetilde \theta \in CI^{(i)}})/L$. To ensure a robust evaluation of the uncertainty performance, various scenarios with different values of measurand $\lambda$ and $a$ are explored. Spectral signatures $X$ are randomly selected based on steel thickness, and the corresponding $\lambda$ is computed. The total counting $ \sum a$  follows a logarithmic-uniform distribution $log(\sum a) \sim U$. The schema of CI evaluation is shown in \autoref{figure:schema_ci}.

The experimental setup for evaluating the LRSR of CIs in this study is summarized as follows:
\begin{itemize}
    \item Two mixtures studied: Bkg with $^{60}$Co and Bkg with four radionuclides $^{60}$Co, $^{137}$Cs, $^{88}$Y, $^{99m}$Tc.
    \item Mixing scenarios: 500 mixing scenarios for Bkg with $^{60}$Co and 1000 scenarios for a mixture of four radionuclides were created with varying counting $a$ and $\lambda$.
    \item MC simulations: for each scenario, 500 spectra were simulated from a Poisson distribution of the theoretical spectrum.
    \item 95.4$\%$ CI calculated by LA and MCMC methods for each of the 500 $\gamma$-spectra simulated for each scenario.
    \item The LRSR value was calculated based on the obtained 500 CIs for each scenario.

\end{itemize}

\begin{figure}[H]
    \centering
      \includegraphics[scale=0.4]{ 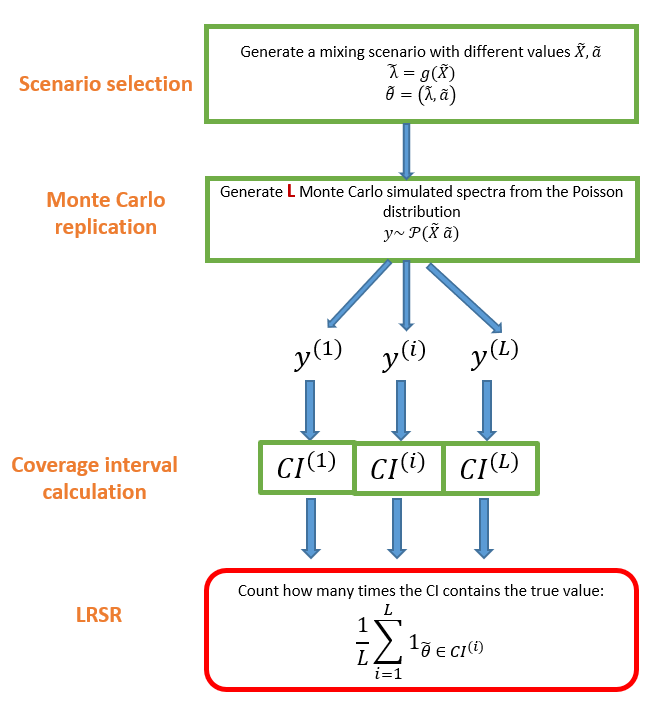}
    \caption{Schema of CI evaluation.}
      \label{figure:schema_ci}
\end{figure}

\section{Results of coverage interval evaluation}
\subsection{Long-run success rate}
\subsubsection{A mixture of Bkg and $^{60}$Co }
\hfill\\

To assess whether the observed LRSR is consistent with the expectation value of 95.4$\%$, an uncertainty bar is defined around the expected value. Each CI either includes the true value or not, which can be modeled as a Bernoulli random variable (1 if it includes the true value, 0 otherwise). Repeating this over $L$ Monte Carlo simulations yields a binomial distribution $B \sim Binom(L,p)$ where $p$ is the expected success rate. The LRSR $R$ is then given by $R=B/L$. Since $L$ is large (500 in this work), the distribution of LRSR can be approximated as $$R \sim Binom(L,p)/L \approx \mathcal N(Lp,Lp(1-p)) /L = \mathcal N(p,\sigma_p^2) $$ where $\sigma_p^2=p(1-p)/L$. Thus, the standard deviation $\sigma_p$ quantifies the variability of the LRSR in different scenarios. To facilitate visualization, different color areas in all figures throughout this study represent the following intervals: within two standard deviations $\sigma_p$, between two and three standard deviations and the rest.

\begin{figure}[H]
    \centering
     \includegraphics[scale=0.3]{ 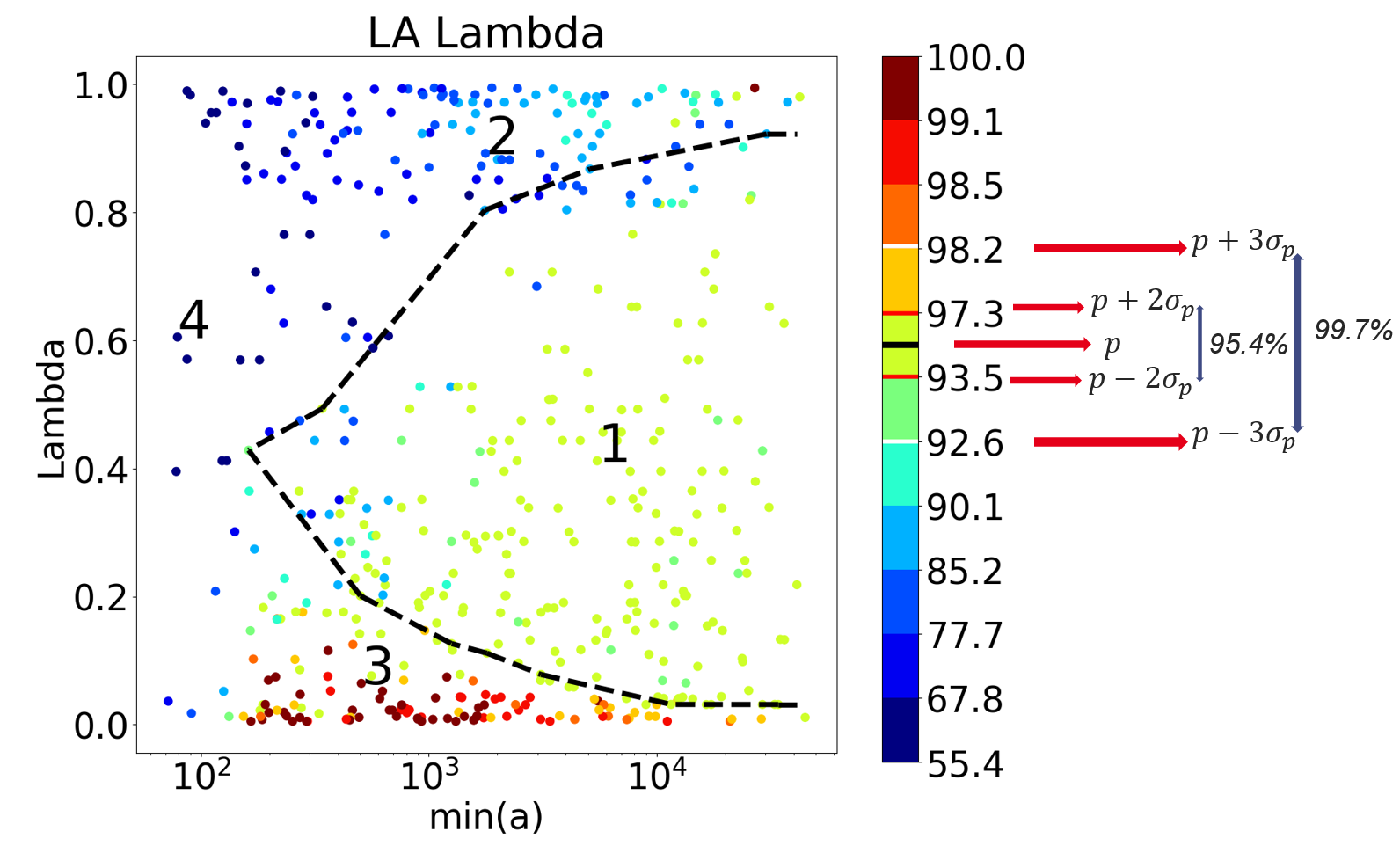}
    \caption{LRSR of 95.4$\%$ CI of the LA method of the latent variable $\lambda$. The x-axis is the minimum counting ($min(a)$) between the Bkg and $^{60}$Co counting. The y-axis represents the $\lambda$ value. The expected value of LRSR in the color bar $p$ is 95.4$\%$. The yellow color corresponds to the interval $[p-2 \sigma _p,p+2 \sigma _p]$, indicating the range where approximately 95.4$\%$ (corresponding to two standard deviations of a Gaussian distribution) of the LRSR values from all scenarios are expected to fall. Orange color represents $[p+2 \sigma _p,p+3 \sigma _p]$. The dashed black curve defines the zone 1  where the scenarios satisfy the criterion discussed above for the LA method.}
    \label{figure:cov_fisher_2_2_exp1}
    \end{figure}

Firstly, a simple mixture of Bkg and $^{60}$Co was studied to evaluate the CI derived from the LA and MCMC methods. In this case, three parameters need to be estimated:  the Bkg counting ($a_1$), $^{60}$Co counting ($a_2$) and the latent variable $\lambda$ characteristic of spectral signatures. \autoref{figure:cov_fisher_2_2_exp1} shows the LRSR obtained by the LA method for $\lambda$ with the expected coverage probability of 95.4$\%$.

As previously mentioned, the LA method may yield imprecise CI when the constraints significantly affect the distribution. The criterion outlined in Section \ref{sec:la} can be employed to verify the applicability of this method. Consequently, all mixing scenarios can be classified into four distinct zones. Zone 1 includes scenarios that meet this criterion, meaning that the constraints on $a$ and $\lambda$ are inactive, resulting in a Gaussian distribution of $\lambda$. In this zone, the LRSR of almost all scenarios falls within the uncertainty bars of the expected value (indicated by the yellow color in \autoref{figure:cov_fisher_2_2_exp1}). Zones 2 and 3 correspond to scenarios where $\lambda$ is close to 0 and 1, related to the maximum and minimum steel thickness. In these cases, the posterior distribution of $\lambda$ is not Gaussian (\autoref{figure:lambda_4zone}), making the CI computed with the LA method imprecise. Zone 4 includes scenarios with very low counting, making it difficult to accurately estimate $\lambda$, which causes a non-Gaussian distribution (\autoref{figure:lambda_4zone}, bottom right). As a result, the LRSR of the LA method in this zone is unsatisfactory.

\begin{figure}[H]
    \centering
      \includegraphics[scale=0.18]{ 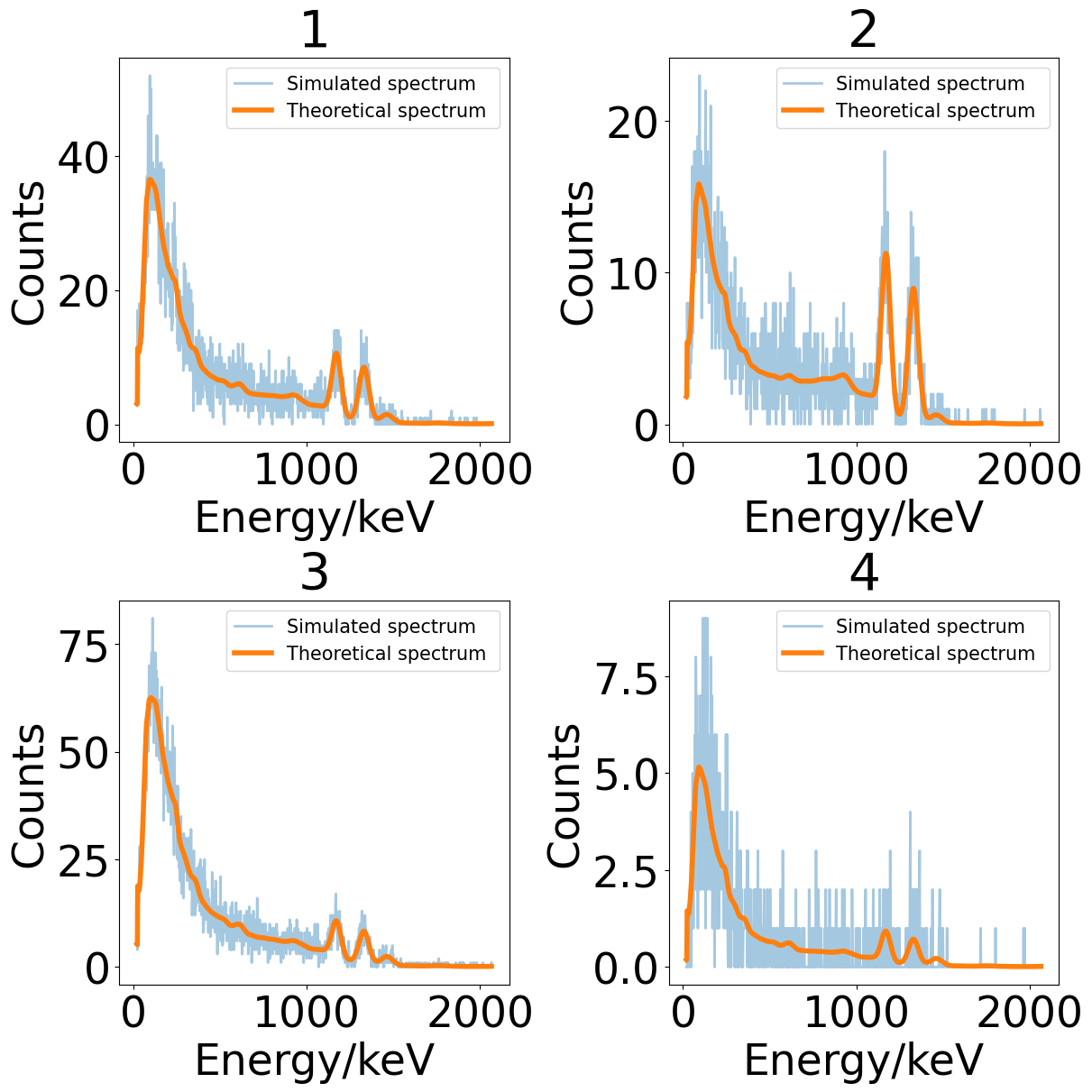}
        \includegraphics[scale=0.18]{ 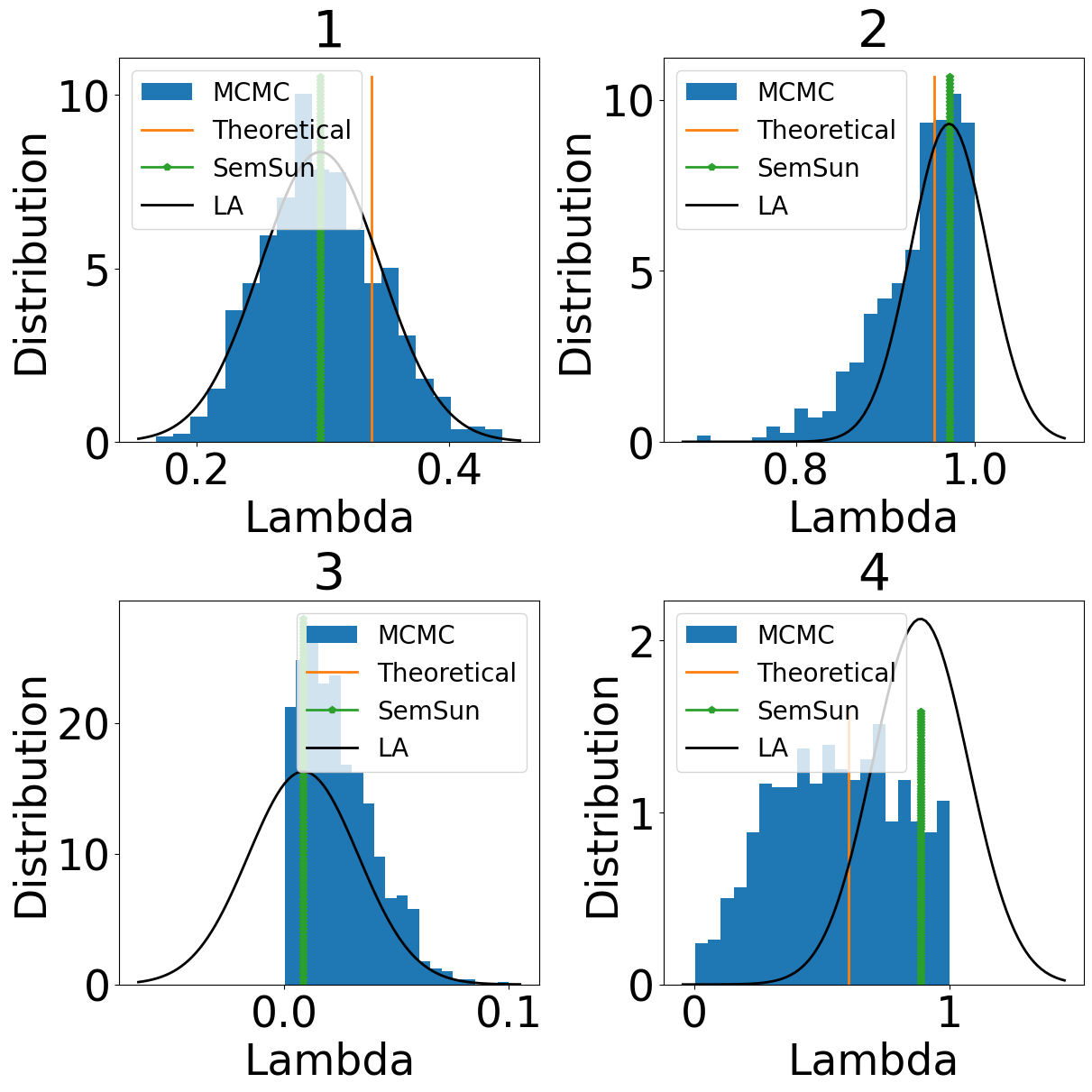}
    \caption{Example of simulated spectra (with Poisson noise) and theoretical spectra (without noise) for fours scenarios corresponding to four zones in \autoref{figure:cov_fisher_2_2_exp1} (left) and associated posterior distribution of the latent variable $\lambda$ (right). They are extracted from the 500 $\gamma$-spectra simulated for each scenario. "MCMC" stands for posterior distribution samples using the MCMC method with 1000 samples. “Theoretical” corresponds to the theoretical value used to create the simulation. "LA" represents the Gaussian distribution obtained by the LA method. "SemSun" stands for the estimation obtained by the SEMSUN algorithm.}
      \label{figure:lambda_4zone}
\end{figure}

Fig. \ref{figure:fisher_mcmc_2_2_exp1} and \ref{figure:fisher_mcmc_2_1_exp1} compare the LRSR of both LA and MCMC methods for the latent variable $\lambda$ and $^{60}$Co. As illustrated, these two methods generally yield similar results of LRSR that align with the expected value (95.4$\%$) when constraints are inactive (zone 1). Certain blue points in zone 1 for the LA method are associated with the cases where the $^{60}$Co counting $a_2$ is very small than Bkg counting $a_1$. In these cases, the high counting of the background in relation to $^{60}$Co yields significantly more noise in the simulated $\gamma$-spectrum, which makes the estimation of $\lambda$ more complicated. This leads to a non-Gaussian distribution of the posterior distribution of $\lambda$ and $a$ for some simulated spectra (see \autoref{figure:multi_mod_exp1}). In scenarios where constraints are active, the MCMC method outperforms the LA method, achieving an LRSR closer to the expected value. Tables \ref{table:la_exp1} and \ref{table:mcmc_exp1} provide additional details on the LRSR for both methods across four example scenarios representing the different zones of LRSR distribution.

It should be noted that in some cases, such as when $\lambda$ is close to 0 or 1, the MCMC result differs slightly from the expected LRSR, as the coverage probability is not exactly the same as LRSR. For instance, the CI calculated using Supplement 1 to the GUM may have a bad LRSR \cite{hall2008evaluating,giaquinto2016examples}. Additionally, when the constraints are applied to the parameters, the LRSR of CI may vary from its coverage probability \cite{marchand2013bayesian}.

\begin{figure}[H]
    \centering
     \includegraphics[scale=0.17]{ 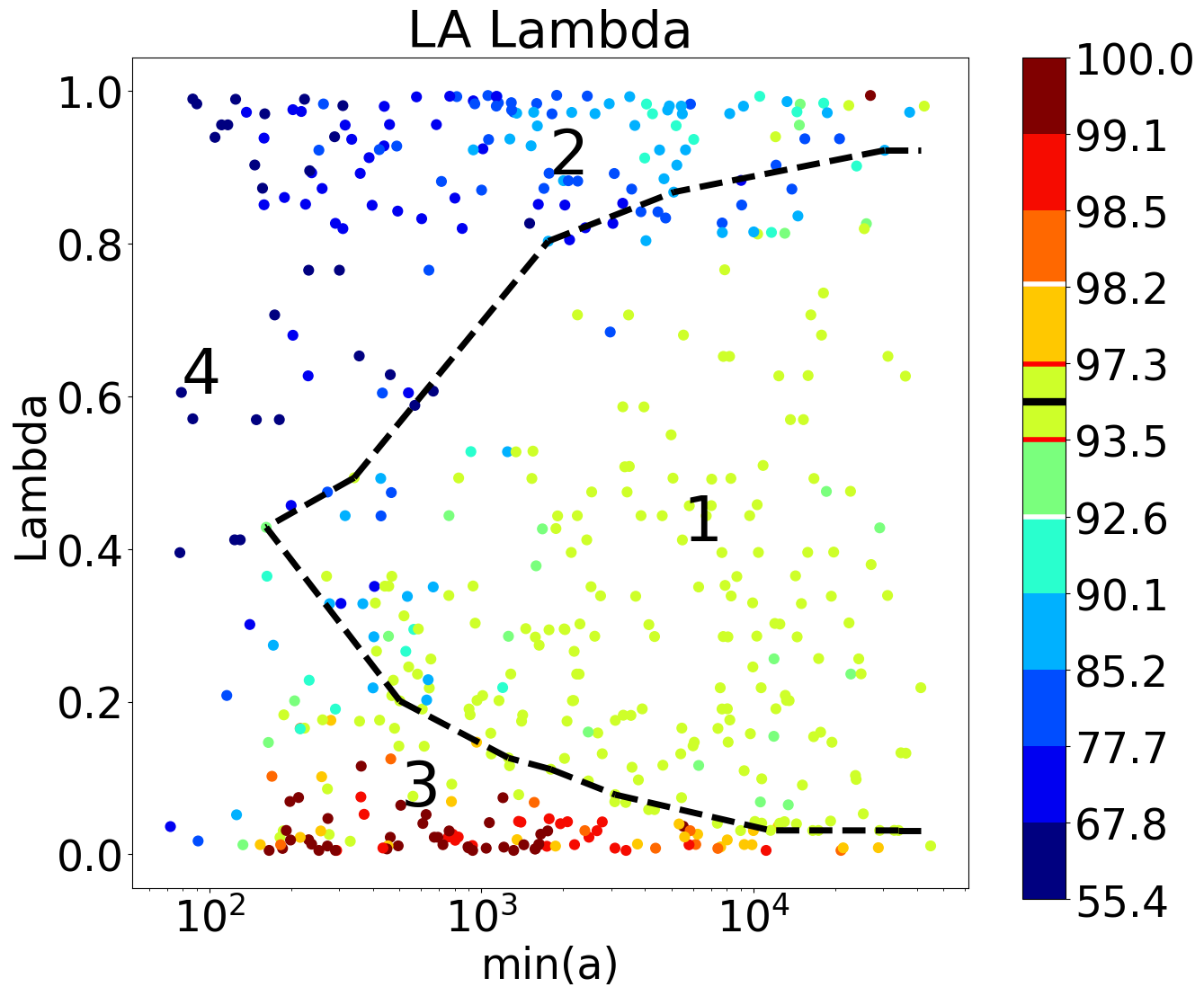}
    \includegraphics[scale=0.17]{ 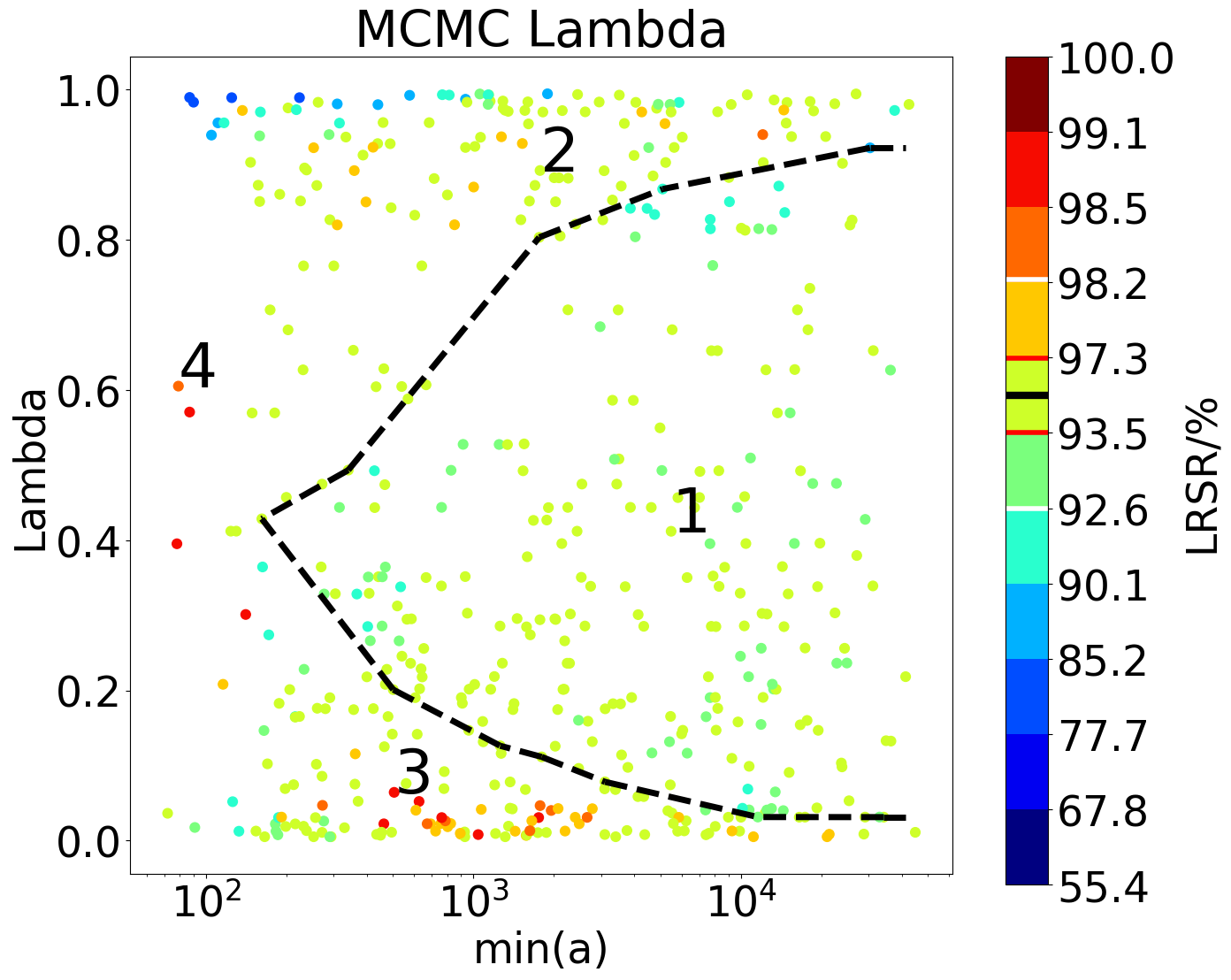}
    \caption{Same as \autoref{figure:cov_fisher_2_2_exp1} for the LA method (left) and the MCMC method (right) for $\lambda$ }
    \label{figure:fisher_mcmc_2_2_exp1}
    \end{figure}

\begin{figure}[H]
    \centering
     \includegraphics[scale=0.17]{ 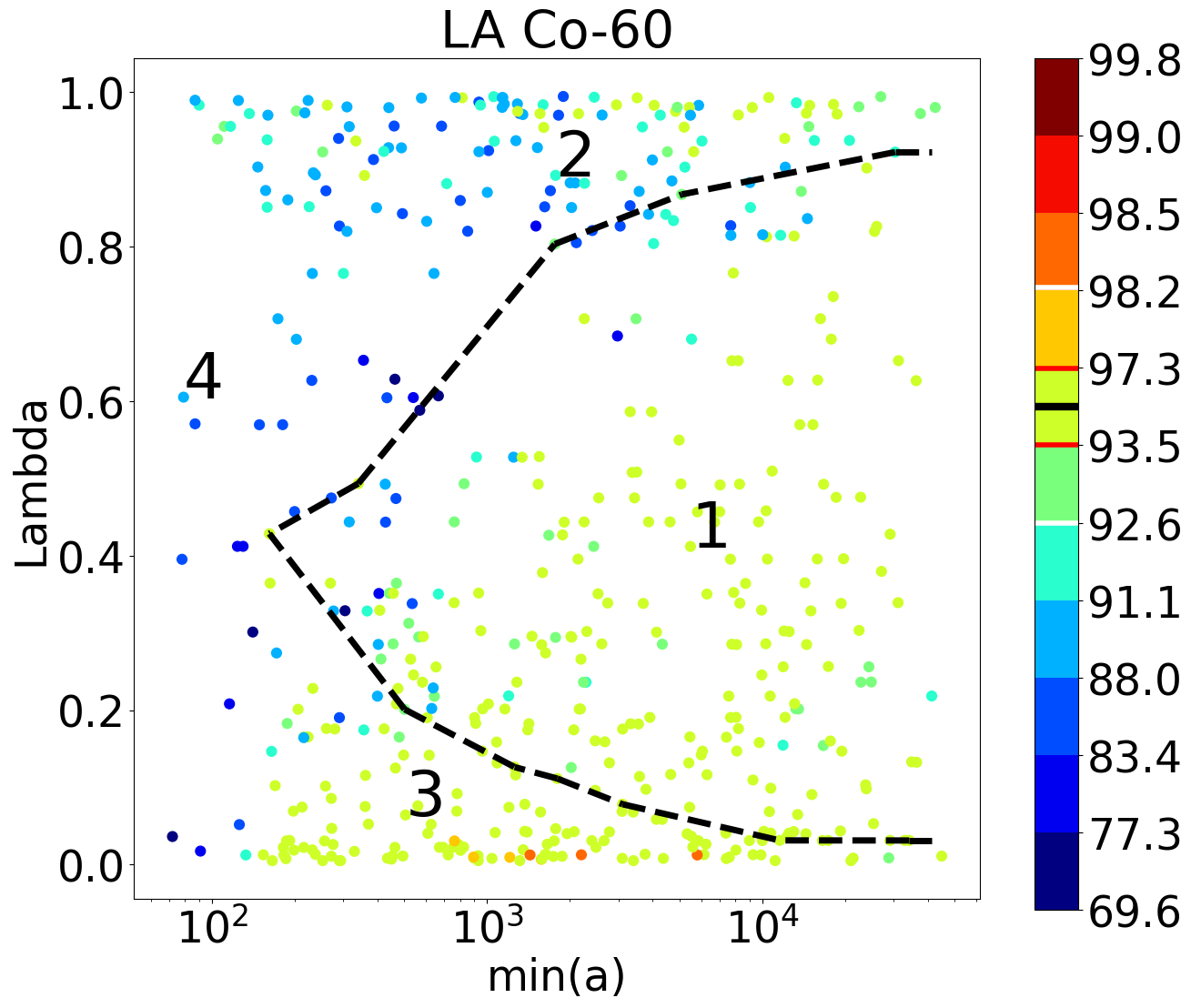}
    \includegraphics[scale=0.17]{ 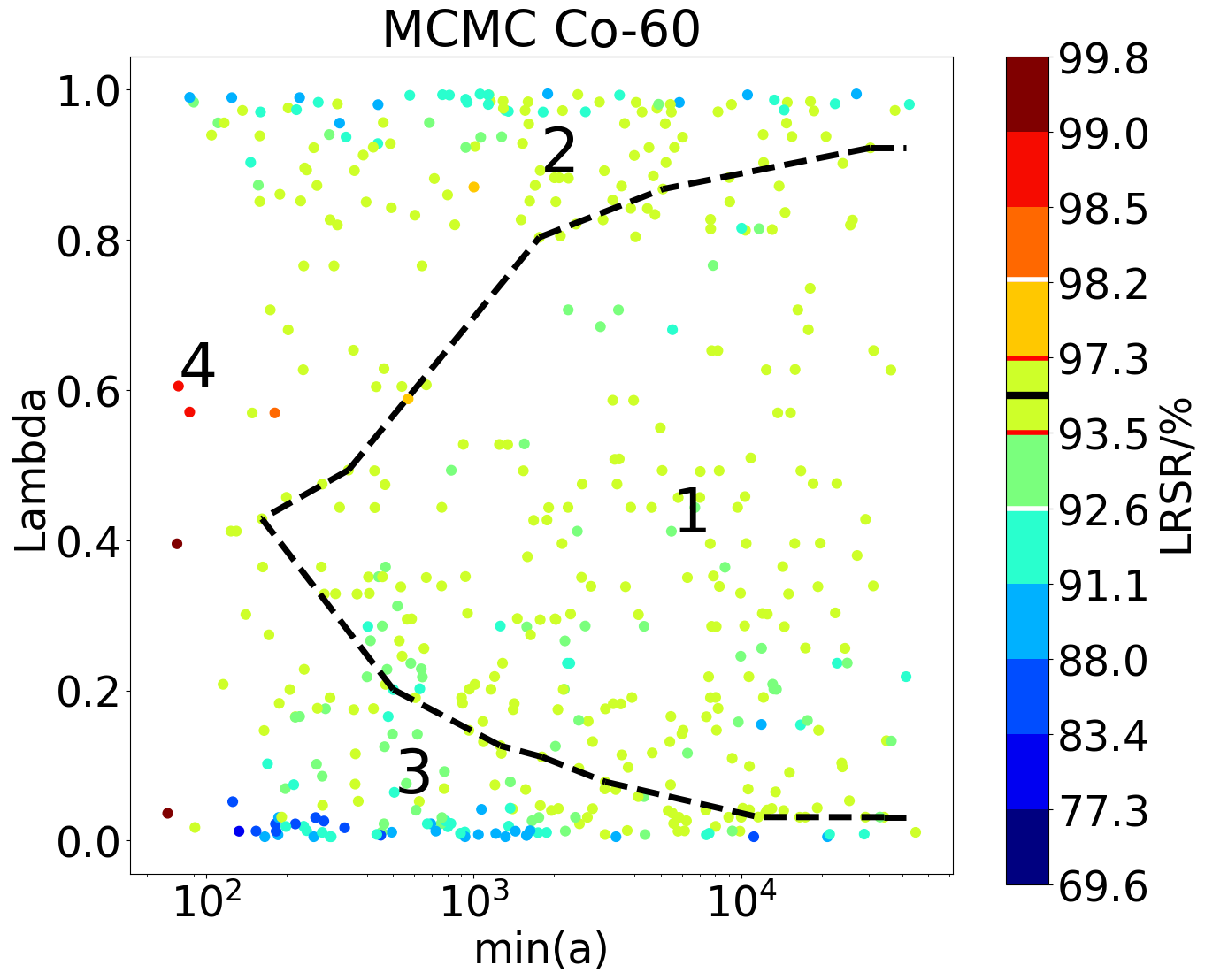}
    \caption{Same as \autoref{figure:fisher_mcmc_2_2_exp1} for $^{60}$Co }
    \label{figure:fisher_mcmc_2_1_exp1}
    \end{figure}

\begin{figure}[H]
    \centering
    \includegraphics[scale=0.2]{ 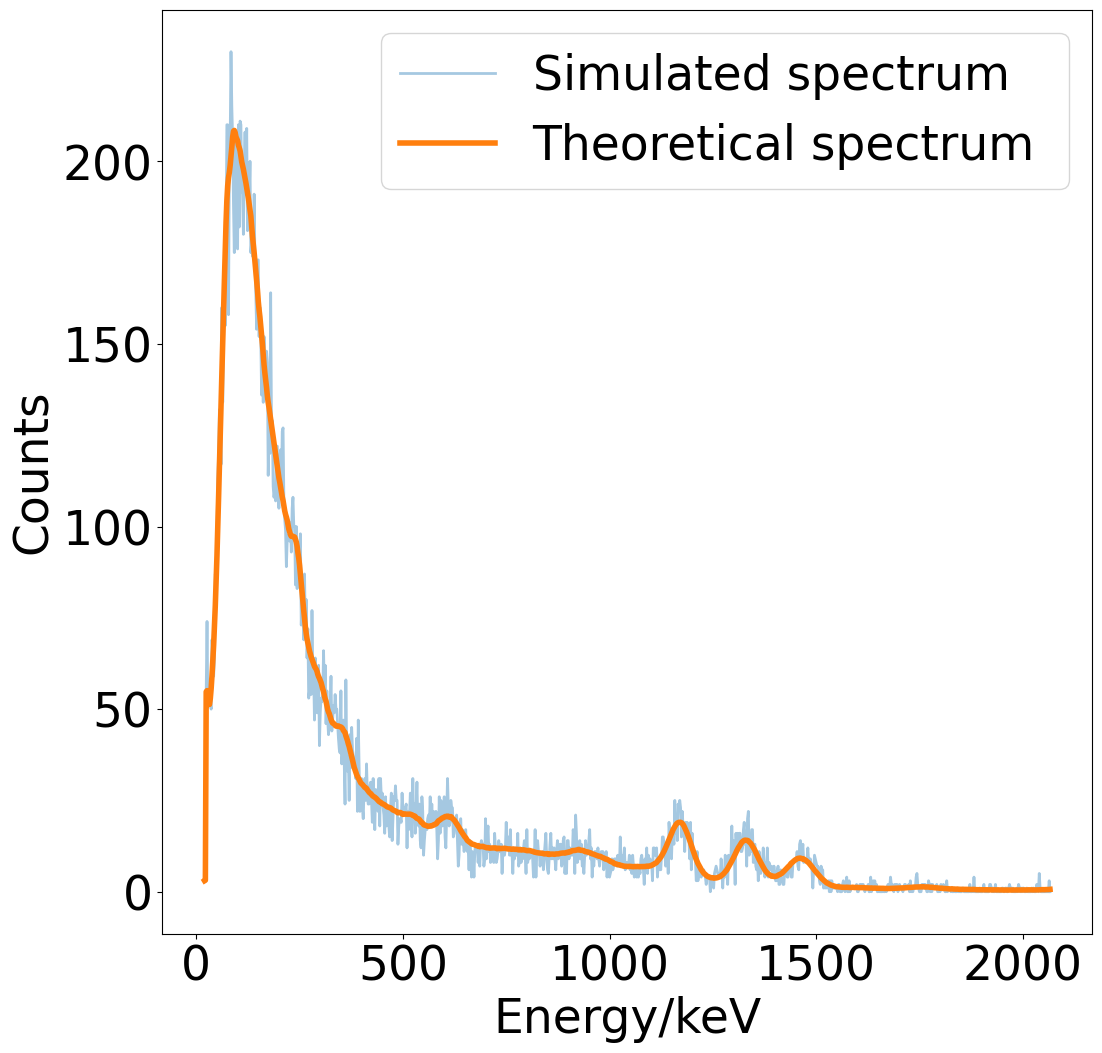}
     \includegraphics[scale=0.18]{ 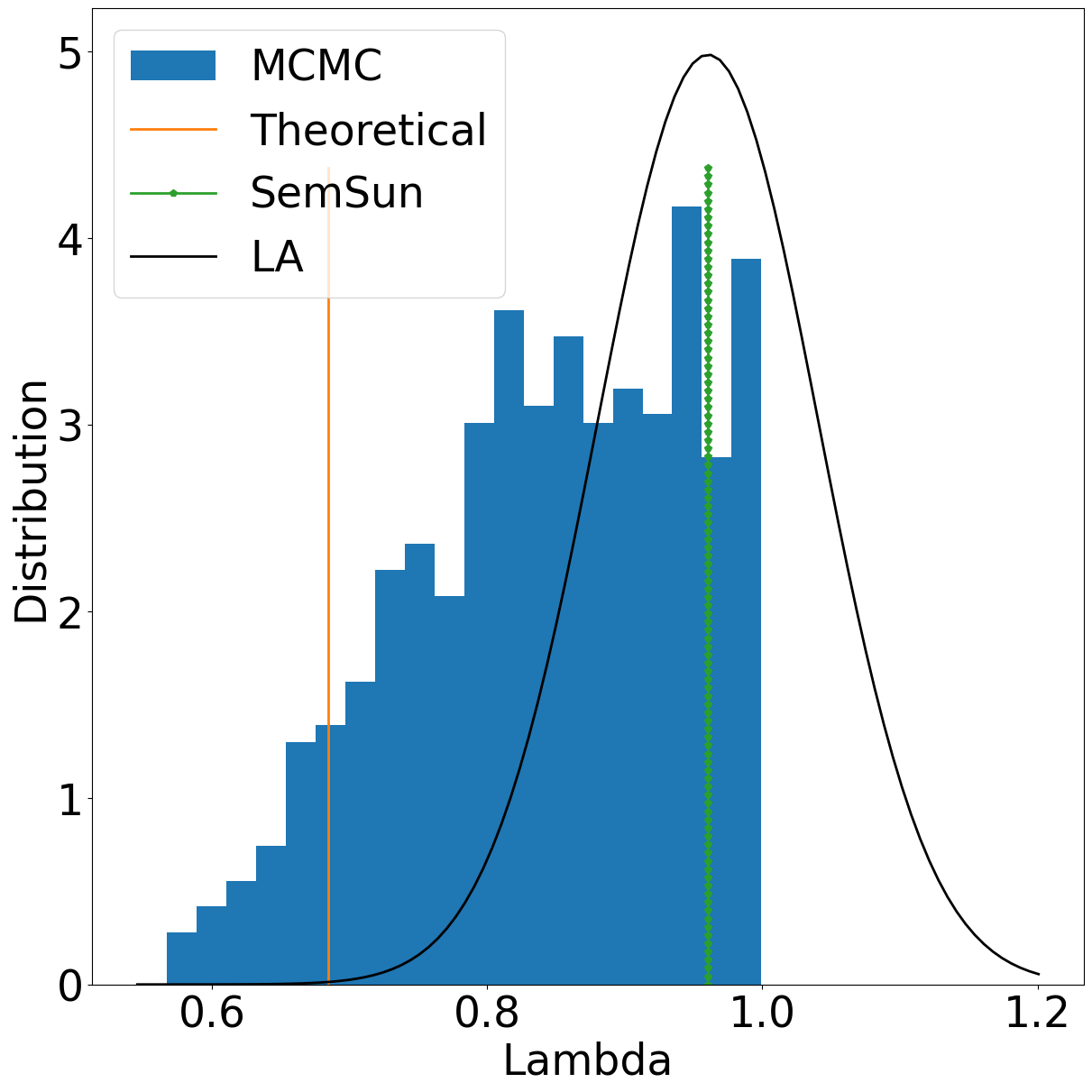}
    \caption{A simulated $\gamma$-spectrum (left) and $\lambda$ distribution (right) for the case of non-Gaussian of the posterior distribution, similar to \autoref{figure:lambda_4zone}.}
         \label{figure:multi_mod_exp1}
    \end{figure}

\begin{table}[H]
\centering
\caption{\label{table:la_exp1} Theoretical values of mixing scenarios and LRSR ($\%$) of 95.4$\%$ CI obtained by the LA method for four examples drawn from each zone shown in \autoref{figure:cov_fisher_2_2_exp1} and \autoref{figure:lambda_4zone}. } 

\begin{tabular}{ccccccc}
\toprule
 &   \multicolumn{3}{c}{Theoretical value}       &  \multicolumn{3}{c}{LRSR($\%$) }          \\
\cmidrule(lr){2-4} \cmidrule(lr){5-7}
Zone &         Bkg &         $^{60}$Co &       $\lambda$ &         Bkg &         $^{60}$Co &       $\lambda$ \\
\midrule
1 &    3782 & 2746 & 0.34 &    94.8  &  95.4  &  95.2  \\
2 &     1608 & 1884 & 0.95 &     91.8  &  93.6  &  88.4  \\
3 &     6346 & 4374 & 0.01 &     95.8  &  96.2  &  98.4  \\
4 &     372 & 109 & 0.61 &        89.6  &  84.6 &  61.8 \\
\bottomrule
\end{tabular}
\end{table}

\begin{table}[H]
\centering
\caption{\label{table:mcmc_exp1} Theoretical values of mixing scenarios and LRSR ($\%$) of 95.4$\%$ CI obtained by the MCMC method for four examples drawn from each zone shown in \autoref{figure:cov_fisher_2_2_exp1} and \autoref{figure:lambda_4zone}. } 
\begin{tabular}{ccccccc}
\toprule
 &   \multicolumn{3}{c}{Theoretical value}       &  \multicolumn{3}{c}{LRSR($\%$) }          \\
\cmidrule(lr){2-4} \cmidrule(lr){5-7}
Zone &         Bkg &         $^{60}$Co &       $\lambda$ &         Bkg &         $^{60}$Co &       $\lambda$ \\
\midrule
1 &    3782 & 2746 & 0.34 &     94.4 &  94.8 &  94.2  \\
2 &     1608 & 1884 & 0.95 &      93.6 &   95.8 &  96.4  \\
3 &     6346 & 4374 & 0.01 &     93.0 &  92.8 &   96.0  \\
4 &     372 & 109 & 0.61 &        97.4 &   98.4 &   97.0 \\
\bottomrule
\end{tabular}
\end{table}

\subsubsection{ A mixture of $^{60}$Co, $^{137}$Cs, $^{88}$Y, $^{99m}$Tc and Bkg}
\hfill\\
A complex mixture involving four radionuclides $^{60}$Co, $^{137}$Cs, $^{88}$Y, $^{99m}$Tc and Bkg was investigated. \autoref{figure:fisher_mcmc_2_5_exp4} shows the LRSR of both methods for the latent variable $\lambda$. Similar to the simple mixture investigated before, the MCMC method generally outperforms the LA method, producing LRSR close to the expected value. When the constraints are inactive (zone 1), the results of both methods are similar except in cases where Bkg counting significantly dominates the other radionuclides, leading to a non-Gaussian distribution similar to the previous mixture (see \autoref{figure:multi_mod}). In such cases, the LA result is poor due to the Gaussian distribution assumption. In contrast, the MCMC result remains close to the expected value since it is derived from the posterior distribution without relying on the Gaussian assumption.

\autoref{figure:fisher_mcmc_2_1_exp4} displays the LRSR for $^{60}$Co counting. Results of other radionuclides such as $^{88}$Y, $^{99m}$Tc and $^{137}$Cs are the same. Generally, the LRSR values of both methods are similar and close to the expected value. This holds true even when $\lambda$ is close to 0 or 1 because, despite the non-Gaussian nature of the $\lambda$ distribution, the associated radionuclide counting often remains Gaussian, thus maintaining the validity of the Gaussian assumption. A low LRSR is only observed in specific cases of low counting, which makes estimating $\lambda$ and $a$ difficult. 

At high statistics ({\it e.g.}, counting around $10^5$), the LRSR of both methods may deviate from the expected value due to the IAE model. Since spectral signatures $X$ are approximated by a generative IAE model, a slight reconstruction error is always present. While this error is usually dominated by Poisson noise and negligible, it becomes significant in very high-statistics cases, affecting the precision of the CI. However, it is important to note that, despite the imperfections of the CI, the difference between the estimated and expected counting remains low.

\begin{figure}[H]
    \centering
     \includegraphics[scale=0.17]{ 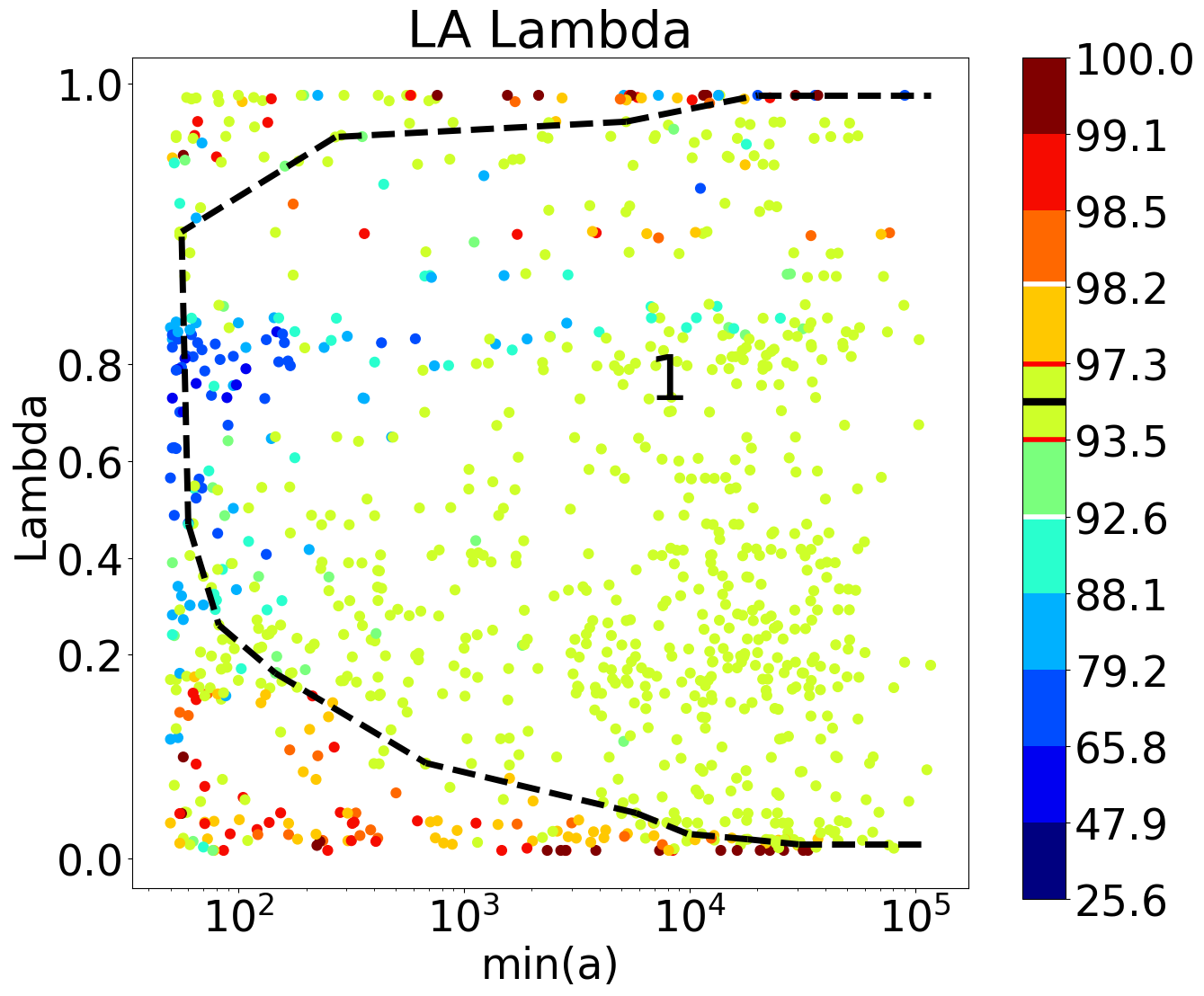}
    \includegraphics[scale=0.17]{ 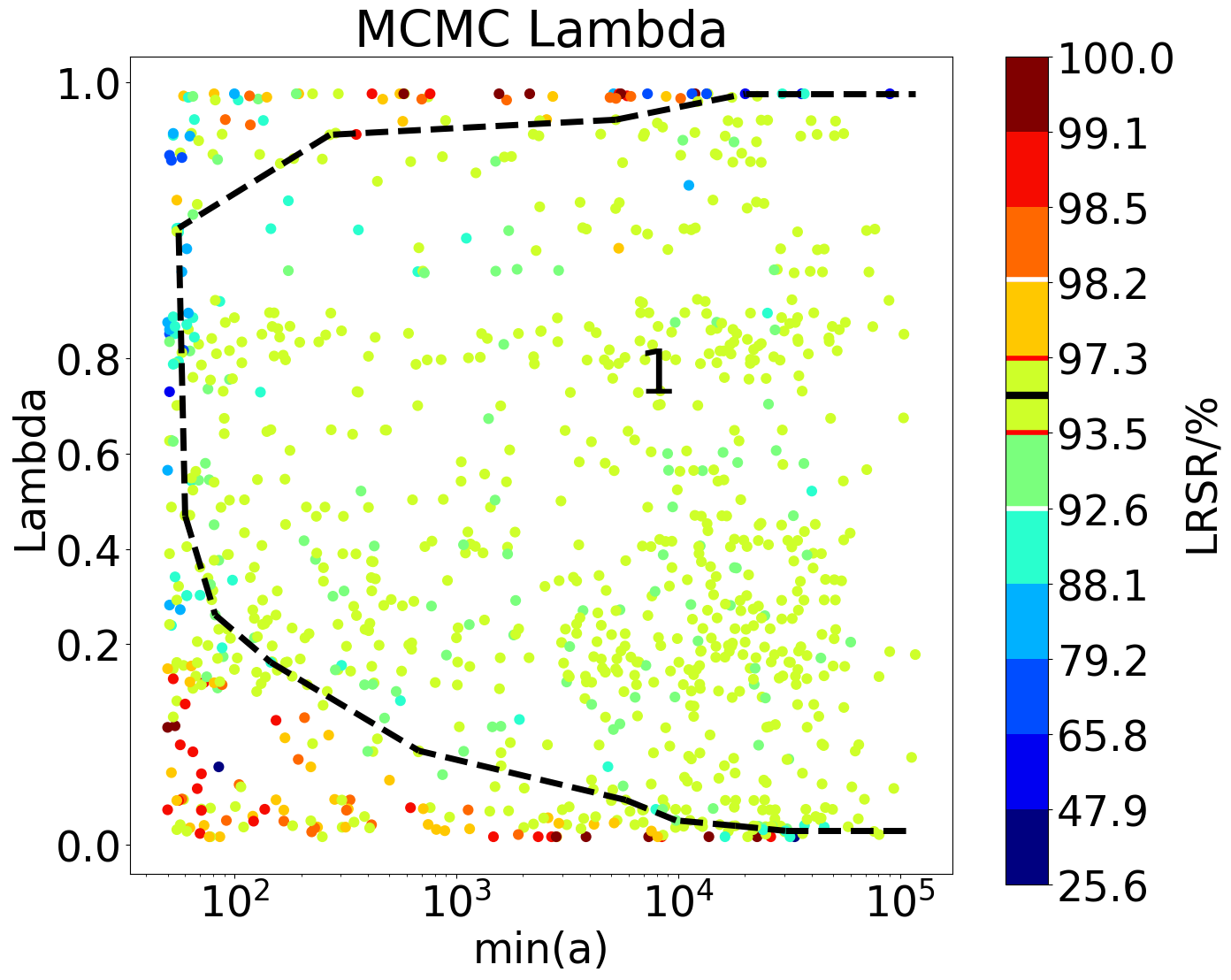}
    \caption{Same as \autoref{figure:fisher_mcmc_2_2_exp1} for the LA method (left) and the MCMC method (right) for $\lambda$. The x-axis is the minimum counting of all radionuclides counting ($min(a)$). The y-axis represents the $\lambda$ value.}
     \label{figure:fisher_mcmc_2_5_exp4}
    \end{figure}
\begin{figure}[H]
    \centering
    \includegraphics[scale=0.2]{ 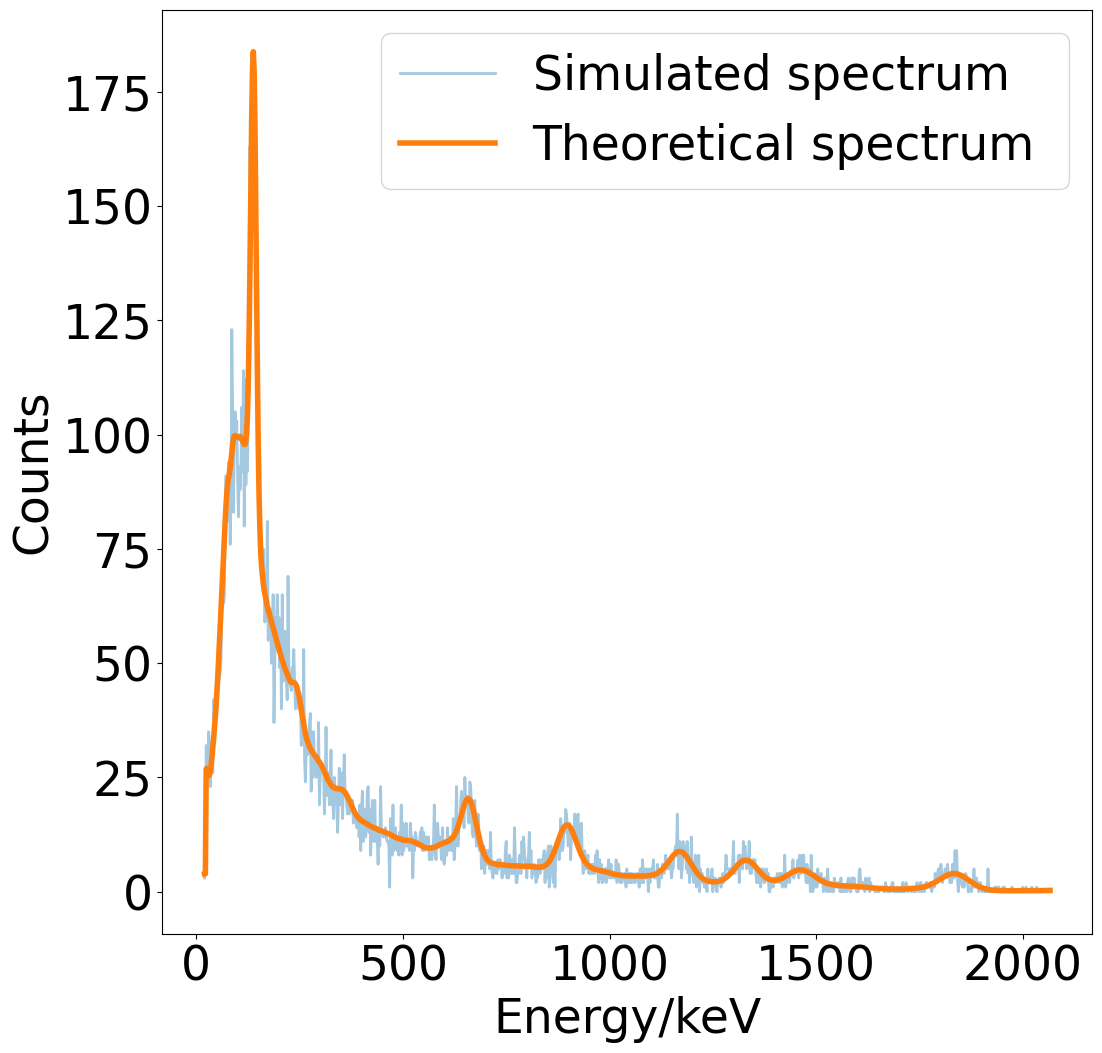}
     \includegraphics[scale=0.18]{ 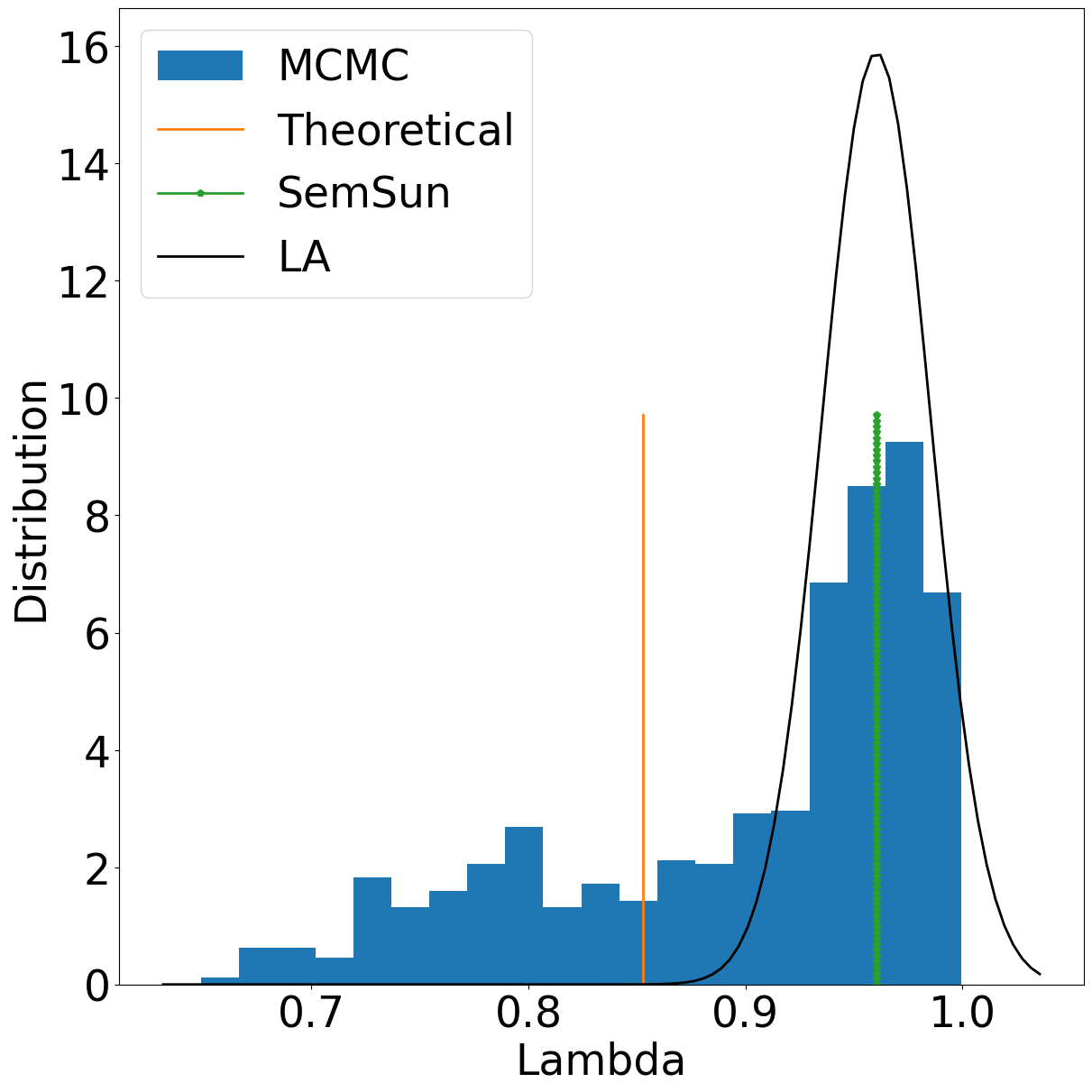}
    \caption{A simulated $\gamma$-spectrum (left) and $\lambda$ distribution (right) for the case of non-Gaussian of the posterior distribution, similar to \autoref{figure:lambda_4zone}.}
         \label{figure:multi_mod}
    \end{figure}

\begin{figure}[H]
    \centering
     \includegraphics[scale=0.17]{ 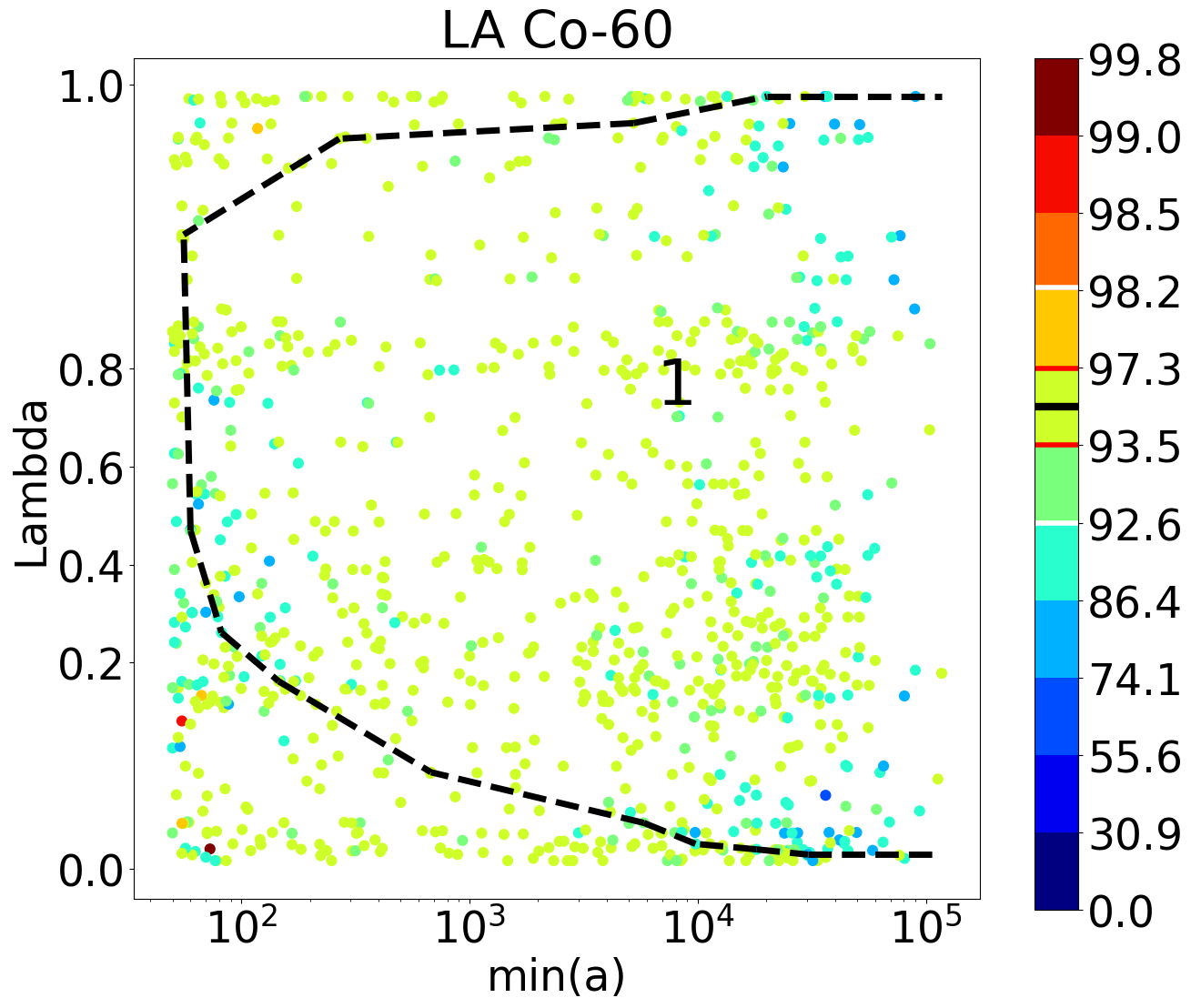}
    \includegraphics[scale=0.17]{ 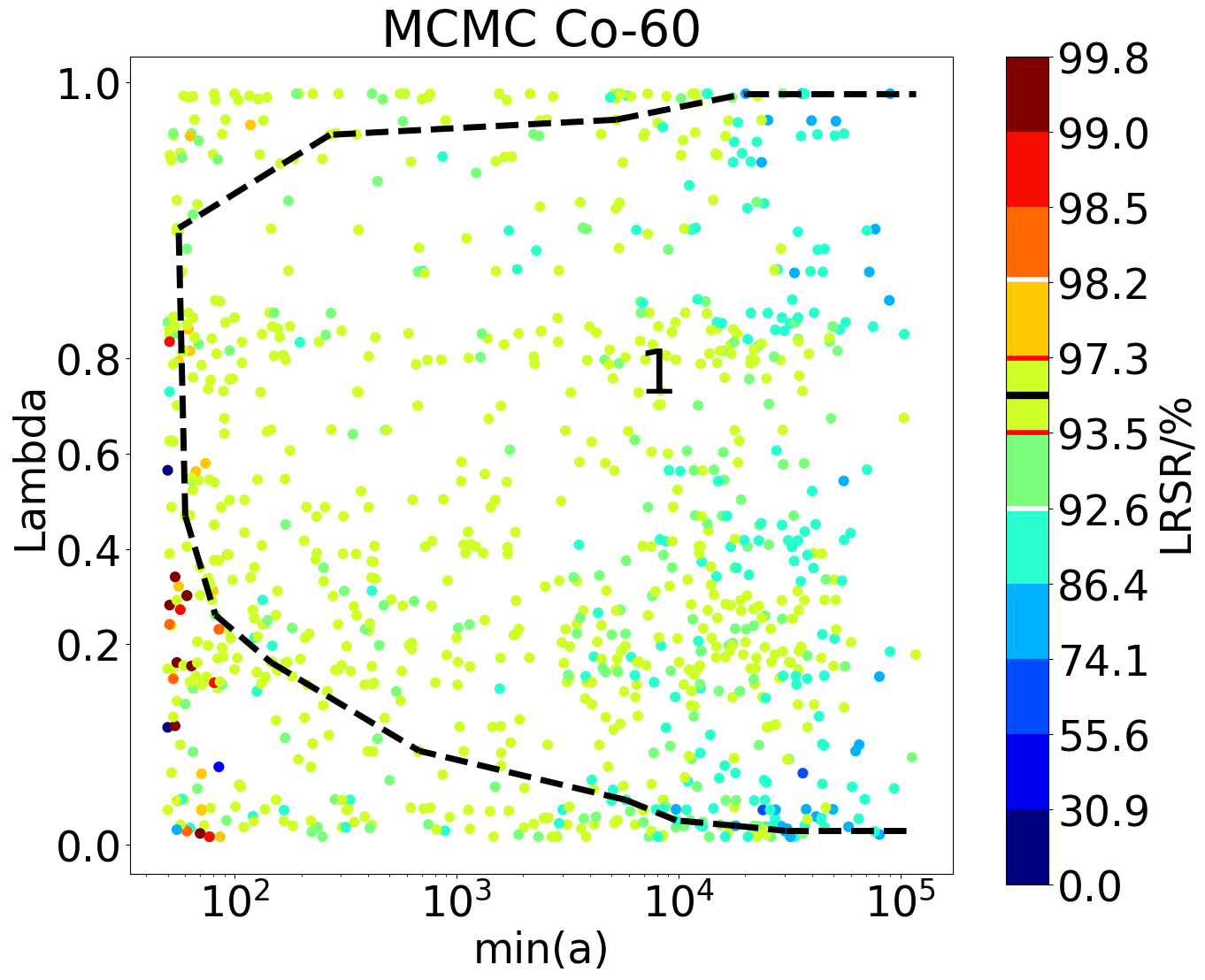}
    \caption{Same as \autoref{figure:fisher_mcmc_2_2_exp1} for the LA method (left) and the MCMC method (right) for $^{60}$Co.}
     \label{figure:fisher_mcmc_2_1_exp4}
    \end{figure}

\subsection{Comparison of the posterior distribution of the LA and MCMC methods}
As demonstrated previously, when the constraints on $\lambda$ and $a$ are inactive, the LRSR of the two methods are similar. This section examines whether the CIs calculated by the two methods are numerically identical. To this end, for each simulated $\gamma$-spectrum, the posterior distributions of $\lambda$ and $a$ derived from MCMC samples are compared with the Gaussian distribution whose mean is obtained from SEMSUN and whose standard deviation is derived from Fisher matrix (see the posterior distribution in \autoref{figure:lambda_4zone}). If these two distributions are similar, the CIs of the two methods should coincide numerically. Consequently, using the CI computed by the LA method can be highly advantageous, as the MCMC method is time-consuming and computationally expensive. Specifically, the calculation time of the LA method is less than 0.1 seconds, while that of the MCMC method is on the order of several minutes.

The total variation distance (TVD) is commonly used in statistics to compare the similarity of two probability distributions. It is defined as half of the L1-norm:
$$\delta(P, Q)=\sup _{A \subset \mathcal{F}}|P(A)-Q(A)|=\frac{1}{2}\|P-Q\|_1=\frac{1}{2} \sum_x|P(x)-Q(x)|$$
The TVD is interpreted as the maximum absolute difference between the probabilities that the two probability distributions attribute to the same event. 

For each simulated spectrum corresponding to each mixing scenario, the TVD was calculated between two histograms: one derived from MCMC samples and the other from the Gaussian distribution of the LA method. The average TVD overall MC simulations for each scenario are shown in \autoref{figure:tv_exp1} for a mixture of Bkg and $^{60}$Co. For the latent variable $\lambda$ and $^{60}$Co, when the constraints are inactive, the posterior distribution can be approximated by a Gaussian distribution, except for a few specific cases discussed earlier, leading to the CI of both methods being numerically similar. However, when the constraints are active, the posterior distribution is no longer Gaussian, resulting in significantly larger TVD values. This outcome indicates that the posterior distributions derived from the two methods differ. The same conclusion holds for the mixture of four radionuclides with Bkg (\autoref{figure:tv_exp4}).

\begin{figure}[H]
    \centering
     \includegraphics[scale=0.17]{ 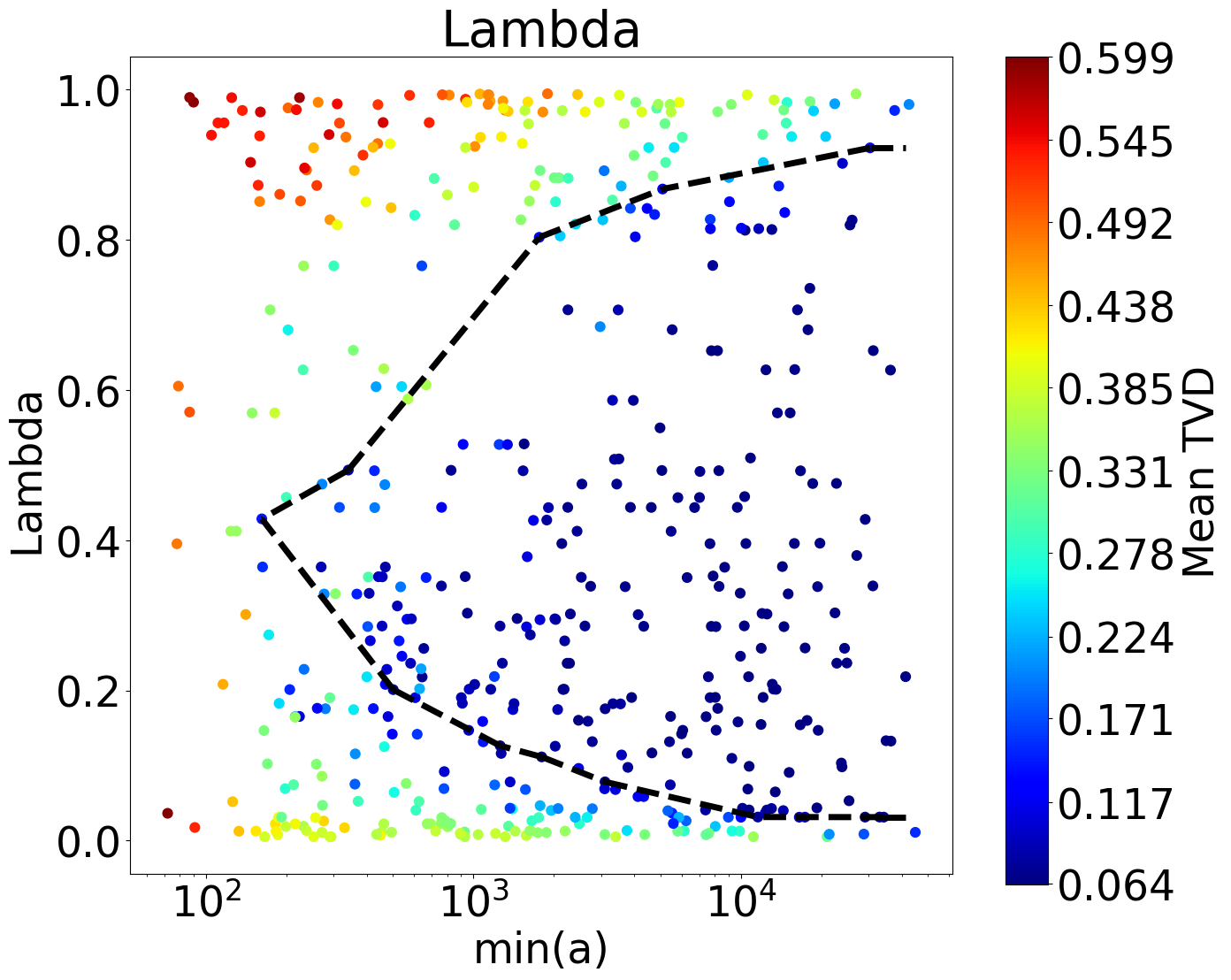}
          \includegraphics[scale=0.17]{ 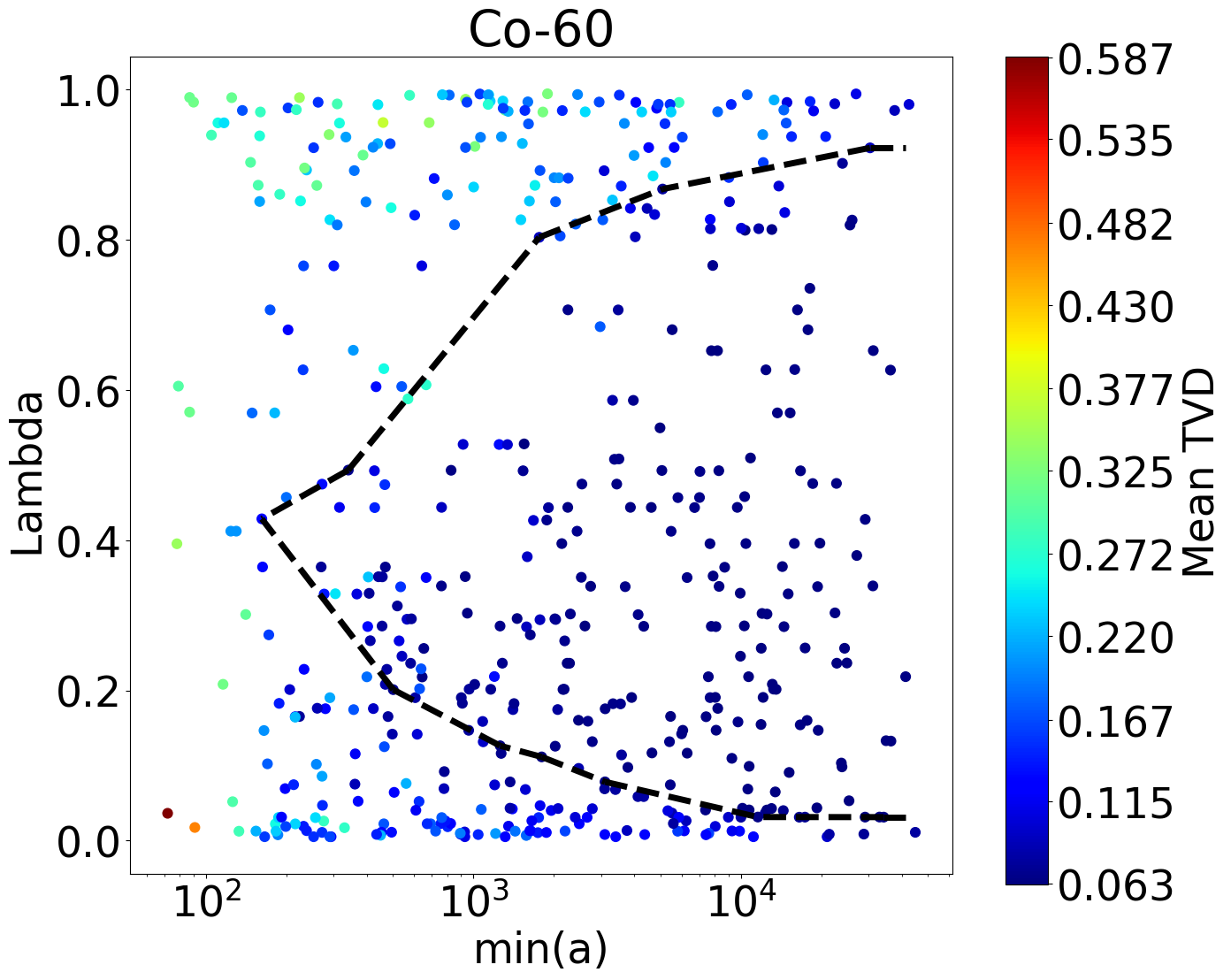}
    \caption{Mean of TVD of all MC simulations for $\lambda$ (left) and  $^{60}$Co (right) for the mixture of Bkg and $^{60}$Co. }
      \label{figure:tv_exp1}
\end{figure}

\begin{figure}[H]
    \centering
     \includegraphics[scale=0.17]{ 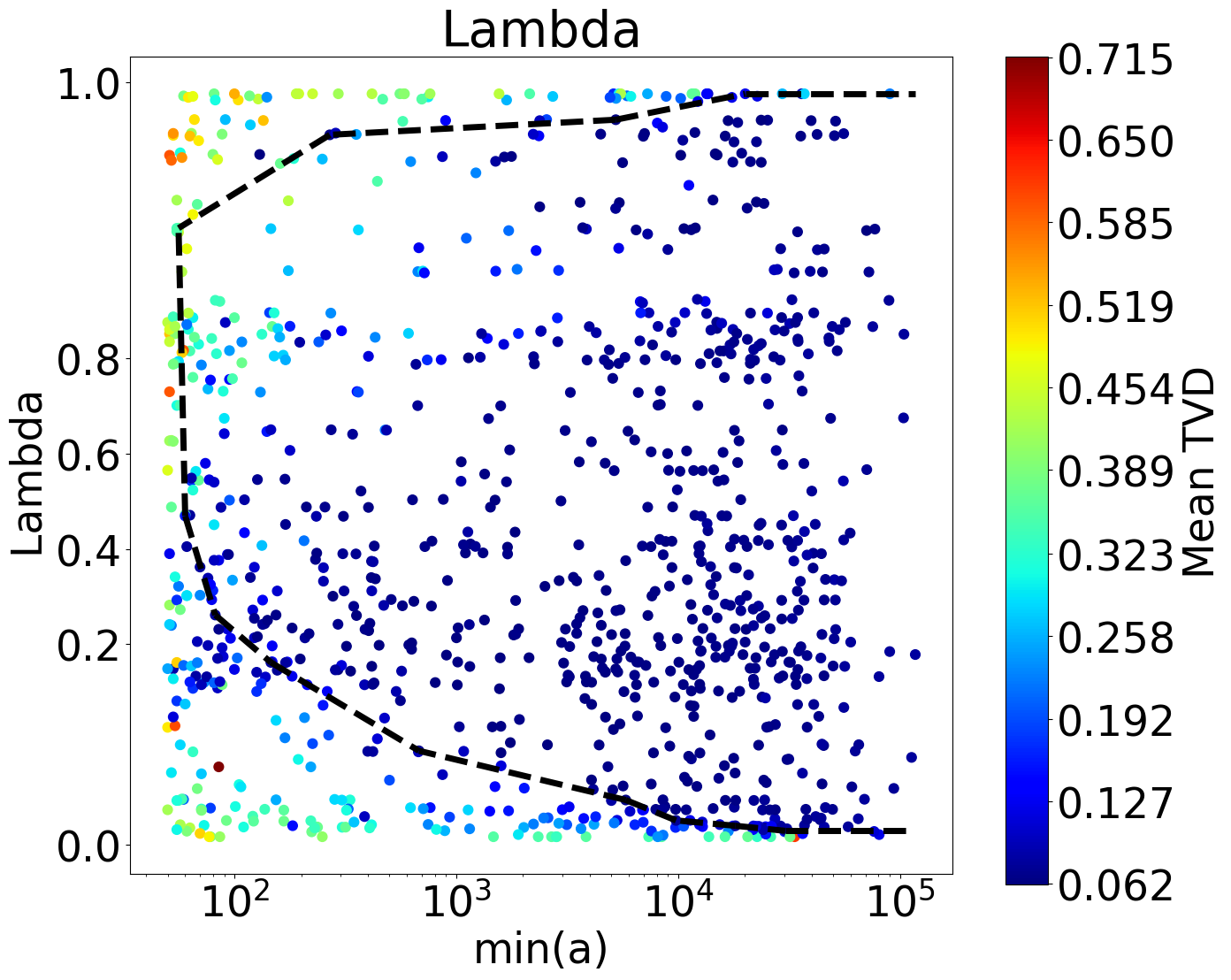}
          \includegraphics[scale=0.17]{ 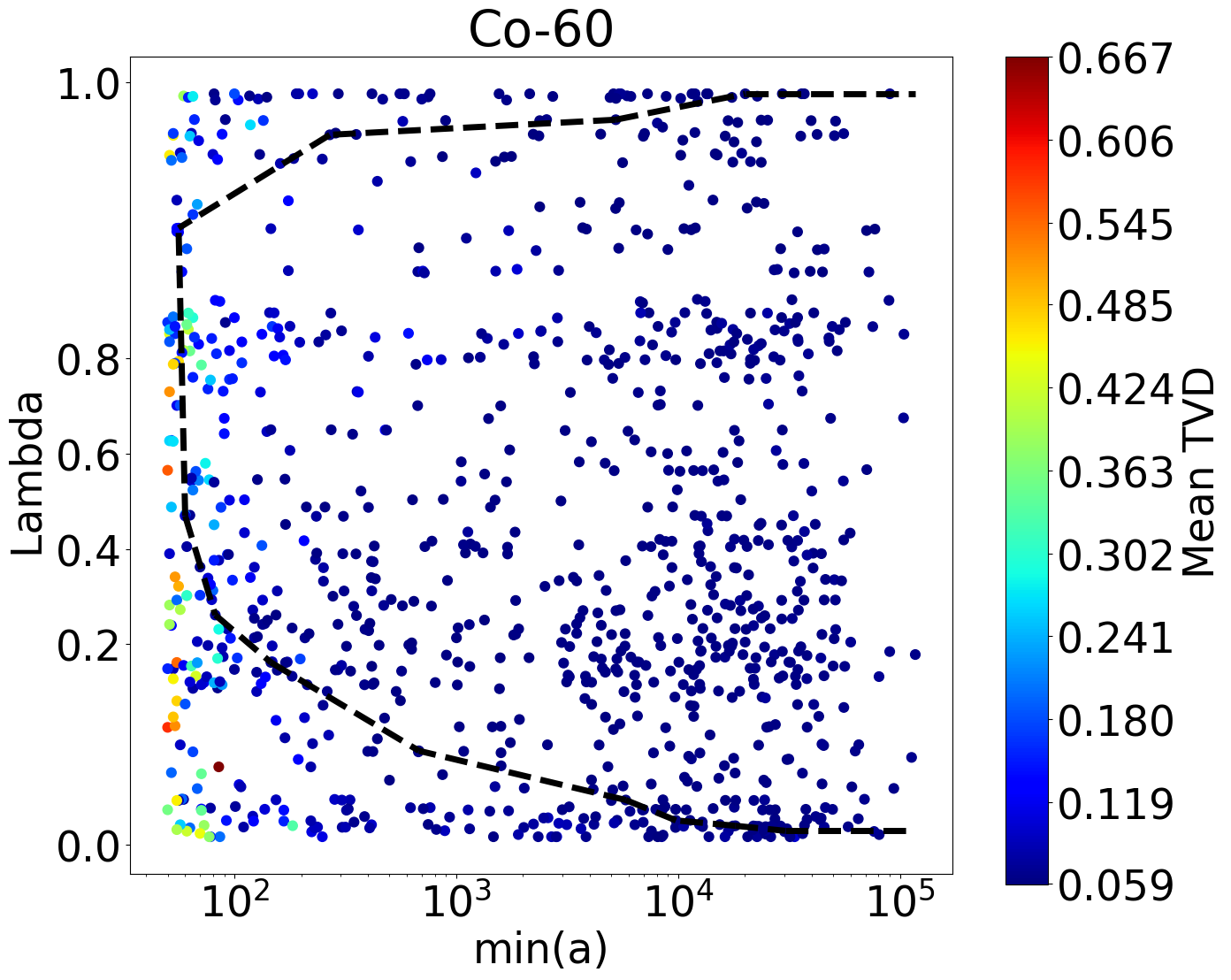}
    \caption{Mean of TVD of all MC simulations for $\lambda$ (left) and $^{60}$Co (right) for the mixture of Bkg and four radionuclides.}
      \label{figure:tv_exp4}
\end{figure}

\section{Conclusion}
In this work, uncertainty quantification of counting and $\lambda$ which characterizes the spectral signatures has been investigated for the hybrid spectral unmixing algorithm SEMSUN in $\gamma$-ray spectrometry with spectral deformation. Given the inverse problem and the constraints involved, two Bayesian methods -LA and MCMC- have been developed to calculate the coverage interval as defined in the GUM. The LA method approximates the posterior distribution by a Gaussian distribution with the mean obtained from the hybrid algorithm SEMSUN, and the standard deviation is derived from Fisher matrix. The MCMC technique samples the posterior distribution by constructing a Markov chain. Two scenarios were investigated: a simple mixture of Bkg and $^{60}$Co and a complex one of four radionuclides $^{60}$Co, $^{137}$Cs, $^{88}$Y, $^{99m}$Tc and Bkg. The performance of both Bayesian methods using the LRSR shows that both methods yield similar results close to the expected success rate of 95.4$\%$ when constraints on the latent variable $\lambda$ related to the spectral signature deformation and counting $a$ are inactive. However, when these constraints are active or Bkg counting significantly dominates other radionuclides, the LA method gives an LRSR different from the expected value due to the non-Gaussian posterior distribution. In such cases, the TVD metric highlights significant differences between the two methods. While the LA method is computationally efficient, the MCMC method offers more robust results when constraints strongly influence the distribution, though at a higher computational cost. Thus, with the criterion based on the activation of constraints and the impact of Bkg counting, we can define cases where both methods give the same coverage interval. In these cases, applying the LA technique is advantageous, while in other cases, the MCMC technique is required to obtain an accurate coverage interval.

\newpage
 \bibliographystyle{elsarticle-num} 
 \bibliography{bibliography}
\appendix

\end{document}